\documentclass[preprint,showpacs,preprintnumbers,amsmath,amssymb]{revtex4-2}
\usepackage{cmap}
\usepackage{graphicx}
\usepackage{graphics}
\usepackage{epsfig}
\usepackage{verbatim}
\usepackage{amssymb}
\usepackage{amsmath}
\usepackage{xcolor}
\usepackage[utf8x]{inputenc}
\usepackage[russian,english]{babel}
\usepackage[figurename=Figure]{caption}
\usepackage{cancel}
\newcommand{\bea}{\begin{eqnarray}}
\newcommand{\beq}{\begin{equation}}
\newcommand{\eea}{\end{eqnarray}}
\newcommand{\eeq}{\end{equation}}
\begin{document}
\title{THz-Driven Floquet Spin Valve-Modulator}
\author{R.G. Nazmitdinov}
\affiliation{Bogoliubov Laboratory of Theoretical Physics, JINR, 141980 Dubna Moscow region, Russia}
\affiliation{Dubna State University, 141982, Dubna Moscow region, Russia}
\affiliation{rashid@theor.jinr.ru}

\begin{abstract}
Within the scattering matrix ($S$-matrix) framework adapted to the high-frequency Floquet--Magnus formalism based on the length gauge, we investigate spin-dependent quantum transport in a 1D quantum ring with an asymmetric
($\pi/2$) configuration of quantum point contacts. We demonstrate the realization of a contactless electromagnetic analog of the Datta--Das spin field-effect transistor with ferromagnetic leads operating at zero static magnetic field. It is shown that an off-resonance terahertz (THz) field enables high-precision switching between spin channels without altering the static parameters of the nanostructure.
Driven by an electronic Vernier effect, the quantum interference in the asymmetric geometry yields a selective spin phase rotator alongside an ultra-high-contrast current-suppression regime (``optical shutter'').
The proposed architecture is fundamentally robust against multiphoton leakage sidebands, offering a thermally immune, high-speed alternative to conventional semiconductor static spin transistors.
\end{abstract}
% PhySH Keywords for APS Submission System:
% Research Areas: Spintronics, Mesoscopic transport, Floquet engineering, Quantum interference.
% Physical Systems: Quantum rings, Quantum point contacts, Semiconductor heterostructures.
% Techniques: Scattering matrix formalism, Floquet-Magnus expansion, Unitary transformations.
%\today
\maketitle

\section{Introduction}

The study of quantum interference in mesoscopic systems traditionally begins with the classic Aharonov-Bohm (AB) effect~\cite{AB}, which demonstrated the physical reality of the electromagnetic vector potential in the phase accumulation of quantum wave functions. An elegant extension of this concept led to the prediction of the Aharonov-Casher (AC) effect~\cite{1}, describing the emergence of a geometric phase in particles with a magnetic moment as they propagate through an electrostatic field. In condensed matter physics, the Rashba-type spin-orbit interaction (SOI) in two-dimensional electron gases has emerged as the direct solid-state analog of this geometric phenomenon~\cite{ras}.

The concept of utilizing SOI for the precise manipulation of spin transport traces back to the seminal work of Datta and Das~\cite{datta}, who proposed the paradigm of the spin field-effect transistor. Within this framework, mesoscopic ring structures stand out as highly promising candidates for quantum interferometers capable of converting spin-dependent phase shifts into a modulation of the total sample conductance. However, as rigorously demonstrated by Meijer et al.~\cite{MMK}, a consistent modeling of such nanodevices within a one-dimensional (1D) approximation necessitates the application of a strictly Hermitian Hamiltonian. Properly accounting for the channel curvature, alongside the mathematically rigorous ordering of the non-commuting momentum and spin operators, is essential for accurately predicting spin-dependent quantum interference effects.

\textcolor{black}{The demand for next-generation, high-speed quantum information processing has driven a concerted search for coherent nanoelectronic components capable of operating within the terahertz (THz) frequency window. While conventional spintronic architectures based on the electrostatic Datta–Das spin field-effect transistor~\cite{datta} stimulated extensive conceptual interest, it is well established that they cannot compete with silicon CMOS in mainstream digital logic~\cite{b1}. As shown by Bandyopadhyay and Cahay~\cite{b1}, static ballistic spin-orbit coupled channels suffer from severe thermal velocity dispersion, which inevitably smears the spin-precession angles and suppresses the modulation contrast at finite temperatures, while operational speeds remain constrained by parasitic gate capacitances. Recently, this has spurred interest in exploring specialized niche applications for spin-manipulation devices, such as quantum state routers and cryogenic interconnects~\cite{b2,b3}.}

To bypass these limitations, all-optical or field-driven ``Floquet engineering'' has emerged as a highly versatile paradigm to manipulate the electronic and spin degrees of freedom on a sub-picosecond timescale without physical contacts~\cite{kib2,she}. By coupling the ballistic carriers to an intense off-resonance high-frequency electromagnetic field, one can achieve a dynamic stabilization of electron states. In this ``radiation-dressed'' regime, the fast time-dependent driving forces collapse into an effective stationary potential that acts as a high-precision, contactless mechanism for quantum phase modulation.

The choice of the material platform plays a decisive role in the physical realization of such light-matter modulated devices. In this context, narrow-gap semiconductor heterostructures based on InGaAs/InAs are highly attractive due to their remarkably high electron mobility, small effective carrier mass, and gate-tunable Rashba spin-orbit coupling~\cite{Nit97,man}. Recent experimental advancements in terahertz spintronics have demonstrated that off-resonance THz driving fields can efficiently induce and modulate transient spin currents in these two-dimensional electron gases (2DEGs) via inverse spin-galvanic and spin-orbit torque mechanisms on a sub-picosecond timescale~\cite{kamr}. However, while macroscopic THz-driven spin transport in unconfined 2DEGs has been widely explored, channeling these dynamic phenomena into structurally confined, mesoscopic coherent devices remains an outstanding challenge, requiring a seamless integration of Floquet engineering with quantum interference.

From a theoretical standpoint, the concept of exploiting high-frequency radiation to tailor the topological and transport landscapes of low-dimensional systems has driven the development of ``Floquet topological insulators'' and dressed-state electronics~\cite{oka,kib3}. When a ballistic system is subjected to an intense, non-resonant high-frequency electromagnetic field, the multichannel scattering matrix can be formulated using a rigorous time-dependent generalization of the Landauer--B\"uttiker framework~\cite{butm,moscal,ihn}. In open quantum networks, such a dynamic drive typically gives rise to multiple sidebands and inelastic channels that can open up non-trivial transport pathways. While previous architectures for Floquet spin filtering often relied on infinite periodic lattices or multi-terminal structures with complex spatial driving profiles, an isolated quantum ring offers a distinctly clean, closed-loop geometry where dynamic phase accumulation can directly compete with geometric and topological phases~\cite{rev}.

Consequently, over the past few decades, the focus of research has shifted toward the analysis and design of efficient spin filters based on Rashba rings (see, e.g., Refs.~\cite{3,bag,jo,ber}), where carrier selection is achieved via the splitting of wave vectors for opposite spin projections. Among numerous approaches, the utilization of high-frequency driving fields for the high-precision control of spin currents in nanoelectronic devices has emerged as a highly promising avenue (see, e.g., Refs.~\cite{she,kozin,5}). The fundamental basis for this framework lies in the concept of "radiation-dressed" matter~\cite{tan}, which leads to the formation of hybrid electron-field quasiparticles—so-called "dressed" electrons. The physical properties of these states differ substantially from those of "bare" carriers, thereby enabling new opportunities for spintronics. Quantum rings provide a compelling platform for implementing dynamic Floquet effects due to their high sensitivity to various geometric phases~\cite{rev}.

To maximize the efficiency of such a phase-driven modulator and enhance its resilience against thermal decoherence, the internal geometry of the interferometer must be carefully engineered. In conventional symmetric Aharonov--Casher rings, multiple internal reflections at the contacts often wash out the conductance modulation, resulting in low-contrast switching profiles that are highly sensitive to thermal energy broadening. To overcome this limitation, one can leverage the mesoscopic analog of the optical Vernier effect by intentionally breaking the structural symmetry of the injector and detector ports~\cite{var1}. The phase mismatch between the individual resonance grids of the unequal arms in an asymmetric ring can suppress unwanted background scattering states while selectively sustaining ultra-sharp, delta-like Airy spikes with a remarkably high effective finesse. When combined with the high dielectric contrast inherent to semiconductor nanostructures embedded in subwavelength environments, this geometric asymmetry serves as a strict prerequisite for achieving a highly robust, trigger-like digital spin-shutter behavior under contactless ponderomotive modulation.

In this work, we combine the concept of geometric phases with the Floquet-Magnus formalism~\cite{f1,f2,f3,f4,f5,f6,oka} to realize an optically controlled spin valve. Utilizing a transition to the dipole approximation, we propose an interferometer model featuring an asymmetric contact configuration ($\pi/2$). We demonstrate that the interplay between the topological Aharonov-Casher phase and the dynamic phase modulation of the wave functions enables both a high-contrast current-suppression regime (acting as an ``optical shutter'') and precise control over the spin precession within the Datta-Das paradigm.

\begin{figure}[h]
\centering
\includegraphics[width=0.65\linewidth]{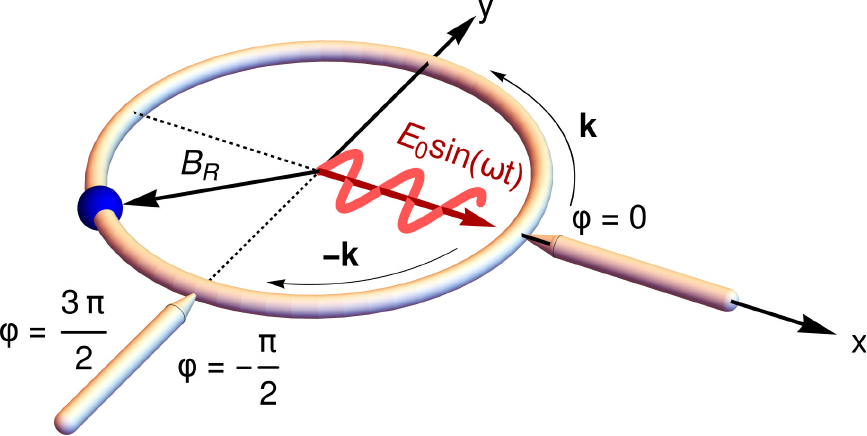}
\caption{(Schematic illustration; not to scale) Sketch of a high-frequency electromagnetic field interacting with an electron propagating inside a
mesoscopic ring. The intrinsic Rashba spin-orbit coupling within the ring generates a wave-vector-dependent effective
magnetic field. The electron spin projections are aligned either parallel or antiparallel to this effective field.
The electron propagates counterclockwise along the upper arm of the ring, moving from the injector quantum point
contact at $\varphi = 0$ to the detector quantum point contact at $\varphi = 3\pi/2$. Conversely, propagation along
the shorter lower arm occurs clockwise, spanning from $\varphi = 0$ to $\varphi = -\pi/2$.}
\label{fig1}
\end{figure}
The defining architectural element of the proposed device is the asymmetric $\pi/2$ configuration of the quantum point contacts (QPCs), which breaks the spatial symmetry inherent to conventional symmetric rings ($L_{\text{up}} = L_{\text{down}} = \pi R$). By setting a strict $3:1$ path length ratio ($L_{\text{up}} = 3\pi R/2$, $L_{\text{down}} = \pi R/2$), the structure operates as a hybrid Mach--Zehnder interferometer coupled with twin nested cavities. This layout leverages the mesoscopic analog of the optical Vernier effect (e.g., Refs.~\cite{var,var1}) within the generalized B\"uttiker scattering framework~\cite{butm,moscal,but}: the phase mismatch between the individual resonance grids of the long and short arms suppresses background scattering states, while selectively sustaining ultra-sharp, delta-like Airy spikes with a remarkably high effective finesse ($\mathcal{F} = 120$). Crucially, as anticipated by the physics of subwavelength slit-guided fields and antenna-coupled modes~\cite{seo}, the tight spatial confinement within the short sub-cavity provides robust spectral isolation ($\hbar \omega \gg \Gamma$), shifting the unwanted inelastic Floquet sidebands far beyond the Fermi energy window. Consequently, this geometric asymmetry serves as a strict prerequisite for achieving a highly robust, trigger-like digital spin-shutter behavior under contactless ponderomotive modulation.

The remainder of this paper is structured as follows. In Sec.~II, we introduce the model Hamiltonian in the G\"oppert--Mayer gauge and derive the field-dressed wave vectors within the Floquet--Magnus framework. Sec.~III details the multichannel $S$-matrix formalism and the exact algebraic factorization of the transmission coefficients. Sec.~IV presents the calculated total conductance and multichannel transport. In Sec.~V, we analyze the signal-to-noise ratio (SNR) and discuss the device performance metrics in relation to realistic InGaAs parameters and antenna-mediated local field enhancement.
Finally, concluding remarks are given in Sec.~VI, while technical details regarding the determinant factorization and the complex pole dynamics
 are relegated to Appendices A and B, respectively.

\section{Hamiltonian and Aharonov-Casher Phase}
\label{hamiltonian}

To describe the spin dynamics of an electron within a one-dimensional (1D) ring of
radius $R$, we employ a strictly Hermitian Hamiltonian~\cite{MMK} augmented by the
interaction with a linearly polarized high-frequency electric field, ${\bf E}(t)=E_0\sin(\omega t)\cdot {\bf e}_x$.
Within the minimal coupling paradigm, $\mathbf{\hat p}\rightarrow \mathbf{\hat p}+e\mathbf{A}(t)$, the
vector potential in the azimuthal direction ${\bf e}_{\varphi}$ reads:
\beq
\label{af}
A_{\varphi}(\varphi, t) = -A_0 \cos(\omega t) \sin\varphi, \quad A_0 = E_0/\omega \,,
\eeq
and the total time-dependent Hamiltonian of the system takes the form:
\begin{equation}
\label{H_init}
H(t) = \hbar\omega_0 \left[ \hat{\ell}_z + \Omega(\varphi, t) \right]^2 +
\hbar\omega_R \left[ \sigma_r \left( \hat{\ell}_z + \Omega(\varphi, t) \right) - \frac{i}{2}\sigma_\varphi \right]\,.
\end{equation}
Here, $\hat{\ell}_z = -i\partial_\varphi$ is the dimensionless orbital angular momentum
operator, $\omega_0 = \hbar/(2m^*R^2)$ defines the spatial quantization scale, $\omega_R = \alpha_R/(\hbar R)$
represents the Rashba spin-orbit frequency, and $\alpha_R$ is the Rashba coupling constant. The dimensionless
field contribution is defined as $\Omega(\varphi, t) = \Omega_0 \cos(\omega t) \sin\varphi$ with the driving
amplitude $\Omega_0 = -e R E_0 / \hbar\omega$.
The spin projection operators onto the local coordinate axes are given
by $\sigma_r(\varphi) = \sigma_x \cos\varphi + \sigma_y \sin\varphi$ and
$\sigma_{\varphi}(\varphi) = -\sigma_x \sin\varphi + \sigma_y \cos\varphi$.

To eliminate the spatial dependence of the Pauli matrices and simplify the spin-orbit interaction
operator, we transform the system into a rotating frame via the unitary transformation
$U = \exp(-i \sigma_z \varphi / 2)$. In this representation, the orbital angular momentum operator
transforms as $U^\dagger \hat{\ell}_z U = -i\partial_\varphi - \sigma_z/2$, while the position-dependent
radial and azimuthal Pauli matrices map onto the stationary matrices $\sigma_x$ and $\sigma_y$, respectively.

When analyzing Floquet systems using the high-frequency Magnus expansion, the conventional choice of
the velocity gauge ($\mathbf{p} \cdot \mathbf{A}$) for the Hamiltonian \eqref{H_init} can lead to a
formal renormalization of the effective electron mass due to the presence of the $\Omega^2$ term in
the kinetic energy operator (see, e.g., Refs.~\cite{she,kozin,5}). Neglecting higher-order corrections
in the high-frequency expansion within this gauge may introduce non-physical artifacts that obscure
subtle physical phenomena. To circumvent these gauge-dependent complications and ensure a transparent
interpretation of the results, we shift to the length gauge (the electric dipole representation)
via the G\"oppert-Mayer unitary transformation (see details in Ref.~\cite{tan1}):
\bea
\label{GM}
V(t,\varphi) = &&\exp\left[\frac{ie}{\hbar}\int A_{\varphi}(\varphi, t) R d{\varphi}\right] = \nonumber\\
=&&\exp\left[ \frac{ie A_0 R}{\hbar} \cos(\omega t) \cos{\varphi} \right].
\eea
The full unitary transformation of the system, defined as $\mathcal{U} = UV$, maps the initial
time-dependent Hamiltonian \eqref{H_init} onto the transformed frame
$\mathcal{H}(t) = \mathcal{U}^\dagger H \mathcal{U} - i\hbar \mathcal{U}^\dagger \partial_t \mathcal{U}$, yielding:
\begin{equation}
\label{h2}
\mathcal{H}(t) = \hat{H}_0 - V_0 \cos\varphi \sin(\omega t).
\end{equation}
Here, $V_0 = e E_0 R$ is the dipole interaction amplitude, and the effective stationary Hamiltonian
$\hat{H}_0$ is defined by the following expression:
\begin{equation}
\label{h3}
\hat{H}_0 = \hbar\omega_0 \left[ \hat{X}^2 + Q \sigma_x (-i\partial_{\varphi})
\right], \quad \hat{X} = -i\partial_\varphi - \frac{\sigma_z}{2}\,,
\end{equation}
where $Q=2m^*\alpha_R R/\hbar^2$.
\textcolor{black}{It is instructive to clarify the status of the Rashba spin-orbit coupling under the applied gauge transformation.
In the minimal-coupling Hamiltonian (2), the vector potential enters both the kinetic energy and the spin-orbit interaction
through the canonical momentum
$\hat{\pi}_{\varphi}=\hat{l}_z+\Omega(\varphi,t)$. Because the G\"oppert--Mayer operator $V(t,\varphi)$ shifts the momentum operator as
$V^{\dagger}(-i\partial_{\varphi})V=-i\partial_\varphi - \Omega(\varphi,t)$ , the gauge transformation simultaneously eliminates
the vector potential from both the orbital kinetic term and the Rashba interaction. Furthermore, the Rashba parameter $\alpha_R$
originates from the static structural inversion asymmetry perpendicular to the 2DEG $(\nabla_z V_{\text{conf}} \mathbf{e}_z)$.
Because the incident THz field is linearly polarized strictly within the plane of the quantum well $\mathbf{E}(t)||\mathbf{e}_x$
with $E_z=0$, the high-frequency drive does not modulate the out-of-plane confinement asymmetry. Consequently, $\alpha_R$
remains an invariant, static material parameter that is unaffected by the dynamic gauge transformation.}

In this representation, the light-matter interaction is mediated
exclusively by the linear dipole term, while the kinetic energy operator fully retains its original form.
This mathematical feature guarantees that the effective carrier mass $m^*$ remains invariant within the
perturbation theory framework. The obtained form \eqref{h2} is optimal for the application of
the Floquet-Magnus formalism, as the time-dependent field drive is entirely decoupled from the
internal kinetic and spin-dependent structures of the system.

In this study, we consider the following realistic material and driving parameters: the electron Fermi
energy $E_F = 1.14$~meV, the driving photon energy $\hbar\omega = 6$~meV, the effective mass $m^* = 0.045m_e$, and the ring
radius $R = 200$~nm. The characteristic spatial quantization scale is $\hbar\omega_0 = \hbar^2/(2m^*R^2) \approx 0.02$~meV.
Consequently, the frequency of the external driving field exceeds the internal orbital frequencies of the system by more
than 300 times ($\omega / \omega_0 \gg 1$, $\hbar\omega \gg E_F$). Within this high-frequency regime, characterized by a
sharp separation of timescales, the electron cannot dynamically follow the instantaneous field oscillations. This justifies
a rigorous transition to the description in terms of field-dressed states. In this representation, the fast temporal
dependencies imposed by the external drive are absorbed into the structure of the modified wave
function, $\psi_{\text{new}}(t) = (UV)^\dagger \psi_{\text{old}}(t)$, via the corresponding phase factors arising
sequentially from both the spatial rotation of the coordinate system and the G\"oppert-Mayer gauge transformation.
This framework allows one to solve the Schr\"odinger equation for $\psi_{\text{new}}(t)$ governed by the transformed
Hamiltonian \eqref{h2}, where the quantum dynamics is determined by an almost stationary effective potential.
According to the Floquet theorem, the solutions within the field-dressed basis take the form:
\begin{equation}
\label{wfn}
\psi_{\text{new}}(\varphi, t) = e^{-i\varepsilon t/\hbar} \Phi_{\text{new}}(\varphi, t),
\quad \Phi_{\text{new}}(\varphi,t) = \Phi_{\text{new}}(\varphi,t+T),
\end{equation}
where $\varepsilon$ denotes the electron quasienergy, and $T = 2\pi/\omega$ is the fundamental period of
the driving field. Crucially, since the combined unitary transformation $\mathcal{U} = UV$ is strictly periodic in time with
the same period $T$, it acts as a Floquet gauge transformation. This mathematically guarantees that the quasienergy
spectrum of the transformed system identically coincides with the physical energy spectrum of the initial
time-dependent Hamiltonian \eqref{H_init}.

At this stage, it is physically instructive to outline the gauge bridge between the continuously dressed states
inside the ring and the transport channels in the external leads. The local G\"oppert-Mayer operator
$V(\varphi, t)$ defined in Eq.~\eqref{GM} essentially accumulates the spatial line integral of the vector potential
from the injection point. Consequently, when evaluating the scattering matrix matching conditions at the boundaries
of the ring, the phase factors generated by the gauge operator $V^\dagger$ at the junction nodes become identically equal
to the time-periodic dynamic phase shifts traditionally obtained via the Peierls substitution framework.
This rigorous mathematical identity ensures that the subsequent transition to discrete Peierls phase factors during
the $S$-matrix factorization remains strictly gauge-invariant and self-consistent.

To determine $\varepsilon$ and analyze the quantum transport dynamics within the field-dressed representation, we
introduce the effective Floquet-Magnus Hamiltonian ${\hat H}_{\text{eff}}$, which is derived via a high-frequency
expansion in terms of the small dimensionless parameter $(\omega_0/\omega) \ll 1$.
To apply the high-frequency expansion, we decompose the time-dependent Hamiltonian \eqref{h2} into its respective
Fourier harmonics as follows:
\bea
\label{fourier_decomp}
&{\cal {\hat H}}(t)={\hat H}_0+{\hat V}_{+1}e^{i\omega t}+
{\hat V}_{-1}e^{-i\omega t},\; {\hat V}_{+1}=-i\frac{{\hat V}_{\text{dipole}}}{2},\;\;\;\;\\
& {\hat V}_{-1}=i\frac{{\hat V}_{\text{dipole}}}{2}\,,\;
{\hat V}_{\text{dipole}} = -V_0 \cos{\varphi}\,,
\eea
where $V_0 = eE_0R$ is the dipole interaction amplitude. According to the high-frequency Floquet-Magnus
expansion, the effective stationary Hamiltonian is evaluated as a series over the driving harmonics:
\begin{align}
\label{hef0}
\;\;\hat{H}_{\text{eff}}={\hat H}_0+\sum_{n>0}\frac{\Big[{\hat V}_n,{\hat V}_{-n}\Big]}{n\hbar\omega}+
\sum_{n>0}\frac{\Big[[{\hat V}_n,{\hat H}_0],{\hat V}_{-n}\Big]}{(n\hbar\omega)^2}+\ldots
\end{align}
The second-order Magnus expansion carried out up to the order of $(\omega_0/\omega)^2$ accurately captures
the dynamic Stark effect and yields the effective spatially inhomogeneous ponderomotive potential to the
energy of the field-dressed electrons:
\begin{equation}
\label{heff}
\hat{H}_{\text{eff}} = \hat{H}_0 + \frac{1}{4\hbar^2\omega^2}\Big[[{\hat V}_{\text{dipole}},
{\hat H}_0],{\hat V}_{\text{dipole}}\Big] = \hat{H}_0 + K(\varphi) \,,
\end{equation}
where the induced time-independent ponderomotive potential $K(\varphi)$ explicitly reads:
\begin{equation}
\label{K_local}
K(\varphi) = \frac{1}{8}\frac{e^2 E_0^2}{m^* \omega^2} \sin^2 \varphi \,.
\end{equation}

The appearance of the positive sign before the ponderomotive term $K(\varphi)$ reflects the standard convention
in Floquet quantum transport, where the time-averaged kinetic energy of the field-induced fast quiver motion
is mapped onto an effective stationary potential barrier. Consequently, for a carrier entering the ring with
 a fixed Fermi energy $E_F$ from the reservoirs, the presence of the field-induced term $K(\varphi) > 0$ effectively manifests as
a local suppression of the available longitudinal kinetic energy ($E_{\text{kin}} = E_F - K(\varphi)$),
 thereby acting as a contactless electromagnetic brake.

Since the spatial modulation amplitude of the induced potential is small compared to the Fermi
energy ($K(\varphi) \ll E_{F}$), the periodic potential relief does not induce significant Bragg
reflection or Floquet band-gap opening near the Fermi level. This justifies the adiabatic phase-accumulation approach,
allowing us to transition to the spatially averaged ponderomotive energy for the analysis of quantum interference effects:
\begin{equation}
\label{K_avg}
\langle K \rangle = \frac{1}{2\pi} \int_0^{2\pi} K(\varphi) d\varphi = \frac{e^2 E_0^2}{16 m^* \omega^2}.
\end{equation}
Here, the total prefactor of $1/16$ arises naturally from the strict algebraic structure of the
second-order Magnus commutator combined with the spatial quantization energy scale $\hbar\omega_0$,
alongside the subsequent geometric averaging of the $\sin^2\varphi$ profile along the perimeter of the
ring (which yields an additional factor of $1/2$). Incorporating the local profile $K(\varphi)$
within the Magnus framework rigorously accounts for the dynamic carrier localization conditions, whereas shifting
to the averaged value $\langle K \rangle$ ensures a self-consistent description of the phase accumulation within
the effective 1D model. For the subsequent analysis of the impact of the incident THz radiation intensity, we define
the driving ponderomotive energy parameter simply as $K \equiv \langle K \rangle$.

\textcolor{black}{Concluding the derivation of the effective description within the high-frequency limit, we note that the application
of the Magnus expansion up to the second order in the small parameter $(\omega_0/\omega) \ll 1$ completely eliminates
the explicit time dependence from the transformed Hamiltonian \eqref{h2}. Taking into account the periodicity of
the Floquet modes, $\Phi_{\text{new}}(\varphi, t) = \Phi_{\text{new}}(\varphi, t+T)$, we expand them into a
Fourier series over the driving harmonics, $\Phi_{\text{new}}(\varphi, t) = \sum_{n=-\infty}^{\infty} \phi_n(\varphi) e^{-in\omega t}$.
This expansion enables the smooth transition from the dynamic Schr\"odinger equation to a stationary eigenvalue problem for the quasienergies $\varepsilon$, defined within each uncoupled photonic subspace index $n$:
\begin{equation}
\label{efin}
\hat{H}_{\text{eff}} \phi_n(\varphi) = (\varepsilon + n\hbar\omega) \phi_n(\varphi).
\end{equation}
Crucially, since the second-order Magnus expansion collapses the explicit time dependence into a purely
stationary effective operator $\hat{H}_{\text{eff}}$, the blocks corresponding to different photon
indices $n$ become structurally decoupled in the high-frequency limit. Here, the effective stationary
Hamiltonian $\hat{H}_{\text{eff}} = \hat{H}_0 + K$ fully incorporates the ponderomotive potential
$K \equiv \langle K \rangle$ induced by the rapid oscillations of the electromagnetic field. While the elastic
transport at the Fermi level is governed primarily by the central $n=0$ block, the presence of the decoupled sidebands
with $n \neq 0$ remains physically vital, as they act as multi-channel dissipation pathways that determine the quantum noise floor of the modulator.
}

The physical significance of the Floquet mode invariance lies in the fact that all "fast" phase fluctuations
within the driving period $T$ are systematically compensated for by the internal structure of the field-dressed
state. Consequently, the phase accumulation as the electron propagates along the arms of the ring is determined
exclusively by the eigenvalues of $\hat{H}_{\text{eff}}$. Thus, the problem of evaluating the quantum interference
landscape reduces to finding the wave vectors $k_n$ from the dispersion relation \eqref{efin} for each
individual photon channel $n$. This approach allows us to interpret the impact of intense radiation as
an effective renormalization of the electron's kinetic energy, which directly manifests as a controlled
shift of the interference fringes upon tuning the external field amplitude $E_0$.

To derive the electron wave vectors for each individual photon channel, we substitute the transformed
momentum operator $\hat{X} = -i\partial_\varphi - \sigma_z/2$ into the stationary eigenvalue problem \eqref{efin}.
In the plane-wave representation, $\phi_n(\varphi) \sim e^{ik_n \varphi}$, the action of the differential operator
simplifies to the algebraic substitution $-i\partial_\varphi \to k_n$. Consequently, the stationary
Hamiltonian $\hat{H}_0$ takes the form of a $2 \times 2$ matrix within the basis of $\sigma_z$ eigenstates:
\begin{equation}
\label{H_matrix}
\hat{H}_0(k_n) = \hbar\omega_0 \begin{pmatrix}
(k_n - 1/2)^2 & Q k_n \\
Q k_n & (k_n + 1/2)^2
\end{pmatrix}.
\end{equation}
Diagonalization of the full effective Hamiltonian matrix, $\hat{H}_{\text{eff}} = \hat{H}_0 + K$, yields the
dispersion relation for each individual spin branch $s = \pm 1$:
\begin{equation}
\label{disp_k}
k_n^2 + \frac{1}{4} + s k_n \sqrt{1 + Q^2} = \frac{\varepsilon + n\hbar\omega - K }{\hbar\omega_0}.
\end{equation}
In an open ring geometry, an electron enters the structure from the reservoirs with a fixed Fermi
energy, $\varepsilon = E_F$ (for the central elastic channel $n = 0$). Since the electronic wave can propagate
both clockwise and counterclockwise, the extraction of the roots naturally introduces a directional
index, $\lambda = \pm 1$, where $\lambda = +1$ corresponds to counterclockwise propagation along the
upper arm and $\lambda = -1$ denotes clockwise propagation along the shorter lower arm (see Fig.~\ref{fig1}).

Solving the quadratic dispersion relation \eqref{disp_k} with respect to the wave vector $k_n$ yields the desired analytical result:
\begin{equation}
\label{kf}
k_{\lambda, s}^{(n)} = -\frac{s}{2} \sqrt{1 + Q^2} + \lambda\sqrt{\frac{E_F + n\hbar\omega - K }{\hbar\omega_0} + \frac{Q^2}{4}}.
\end{equation}
Remarkably, we obtain a set of four distinct field-dressed wave vectors for a given energy. To capture
the full manifold of transmission eigenstates, we utilize the substitution $k \rightarrow k_{\lambda, s}^{(n)}$.
In what follows, we introduce several compact and computationally convenient definitions:
\begin{align}
\label{A}
A &= \sqrt{1+Q^2} \quad \Longrightarrow \quad Q^2=A^2-1\,, \\
\label{C}
C_n &= \sqrt{A^2-1+\frac{4(E_F-K+n\hbar\omega)}{\hbar\omega_0}}\,, \\
\label{kls}
k_{\lambda,s}^{(n)} &= \frac{1}{2}(\lambda C_n - sA).
\end{align}
Equation \eqref{kls} defines the field-dressed wave vectors responsible for the phase accumulation of the electron
as it propagates along the arms of the ring. The field-induced ponderomotive contribution $K$ effectively suppresses
the kinetic energy of the charge carriers, which is algebraically equivalent to a shift in the wave number.
This resulting formulation allows us to decouple the dynamics of the dressed electrons into two primary physical driving factors:
\begin{itemize}
\item {\bf Parameter $C_n$ (Kinetics + Light):} This parameter governs the longitudinal propagation of the carriers.
It monotonically decreases upon increasing the laser intensity because of the growth of the ponderomotive barrier $K$.
Thus, the high-frequency radiation acts as an \textit{effective electromagnetic brake}, dynamically shifting the
quantum resonance conditions inside the ring. Crucially, this term is strictly independent of the spin projection $s$.

\item {\bf Parameter $A$ (Topology + Rashba):} This parameter dictates the spin-orbit splitting. It shifts the
momentum upward for one spin projection and downward for the opposite one. It is this term that inherently establishes
the phase mismatch between the $\uparrow$ and $\downarrow$ components, enabling the ring architecture to function as
a coherent \textit{spin phase modulator}.
\end{itemize}
Thus, the dispersion relation \eqref{kls} establishes the theoretical foundation for the quantitative description of
the laser-induced modulation of the quantum interference current, where the external high-frequency field acts as a
high-precision, contactless, optically driven spin shutter.

To provide a transparent physical visualization of these quantum interference phenomena, let us schematically outline
the evolution of the wave function. Suppose that at the injection node ($\varphi = 0$), the incoming electron is
described by a constant spinor $\chi_s$ aligned with the direction of the local effective Rashba spin-orbit magnetic
field. Upon propagating along either the upper or lower arm of the ring (indexed by $\lambda = +1$
and $\lambda = -1$, respectively), the emergent electronic wave function at the detector drain
junction ($\Psi_\lambda$) acquires the following factorized form:
\begin{equation}
\Psi_{\lambda,s}^{(n)} (t) = \mathcal{A}_\lambda
\underbrace{
\exp \Big(i \frac{\sigma_z}{2}
\varphi_{\lambda}\Big)}_{\text{Topology}}
\cdot
\underbrace{e^{i k_{\lambda,s}^{(n)}\varphi_{\lambda}}}_{\text{Kinematics}} \cdot
\underbrace{\exp\big(- i z \cos(\omega t) \big)}_{\text{Dynamics}}
\chi_s,
\label{stoke}
\end{equation}
where $\mathcal{A}_\lambda$ is the transmission amplitude along the respective arm, $\varphi_\lambda$ denotes
the net azimuthal angle of the geometric path, and $k_{\lambda,s}^{(n)}$ represents the effective field-dressed
wave vector defined by Eq.~\eqref{kls}.

Equation \eqref{stoke} explicitly decouples three distinct physical mechanisms:
\begin{enumerate}
    \item \textbf{Topological Phase:} The first prefactor describes the inverse transformation of
    the spinor from the rotating frame back to the laboratory coordinate system. Given that the geometric
    path length difference between the two arms equals exactly $2\pi$ ($\varphi_{\lambda=+1} = 3\pi/2$
    and $\varphi_{\lambda=-1} = -\pi/2$), the fundamental rotation properties of a spin-1/2 spinor strictly
    dictate a relative minus sign: $\exp(i\sigma_z \varphi_{\lambda=+1}/2) = -\exp(i\sigma_z \varphi_{\lambda=-1}/2)$.
    This topological phase shift serves as the primary mechanism ensuring destructive quantum interference in the
    stationary, unperturbed regime.

    \item \textbf{Kinematic Phase:} This term represents the static phase accumulated by the electron
    within the $n$-th channel, defined as $\Phi_{\lambda,s}^{(n)}=k_{\lambda,s}^{(n)}\cdot \varphi_{\lambda}$,
    during its ballistic propagation along:

    a) the upper long arm ($\varphi:0\rightarrow 3\pi/2$):
    \begin{align}
    \label{st1}
    \Phi_{\lambda=+1,s}^{(n)}= \frac{3\pi}{4}C_n - s\frac{3\pi}{4}A \,,
    \end{align}

    b) the lower short arm ($\varphi:0\rightarrow -\pi/2$):
    \begin{align}
    \label{st2}
    \Phi_{\lambda=-1,s}^{(n)}=\frac{\pi}{4}C_n + s\frac{\pi}{4}A \,,
    \end{align}
    where the parameter $A$ is defined by Eq.~\eqref{A}, and the channel-specific parameter $C_n$ is defined by Eq.~\eqref{C}.

  \item
  \textbf{Dynamic Phase (Floquet Dressing):} When an electron propagates through the ring in the
  presence of an alternating electromagnetic field, its wave function acquires an additional time-dependent
  gauge phase, which is conceptually equivalent to the Peierls substitution framework. Specifically, the
  vector potential $\mathbf{A}(t)=\mathbf{A}_0 \cos{(\omega t)}$ contributes the following dynamic phase factor:
\beq
\theta_{\text{dyn}}(t)=\frac{e}{\hbar}\int \mathbf{A}\cdot d\mathbf{l}\,.
\eeq
For a driving field linearly polarized along the $x$-axis, the line element integration yields
$\mathbf{A}\cdot d\mathbf{l} = -A_0\cos{(\omega t)} R\sin{\varphi} d{\varphi}$, which results in
the following phase accumulations along the respective trajectories:
\begin{align}
\label{df1}
\theta_{\text{dyn}, \lambda=+1}(t)& = -z\cos{(\omega t)}\int_0^{3\pi/2} \sin \varphi d\varphi = -z\cos{(\omega t)}\,,\\
\label{df2}
\theta_{\text{dyn}, \lambda=-1}(t)& = -z\cos{(\omega t)}\int_0^{-\pi/2} \sin \varphi d\varphi = -z\cos{(\omega t)}\,,
\end{align}
where the dimensionless driving amplitude parameter $z$ is explicitly defined as:
\beq
z=\frac{eRA_0}{\hbar}\,.
\eeq
\end{enumerate}

It is highly instructive to recognize that the identity
$\theta_{\text{dyn}, \lambda=+1}(t) = \theta_{\text{dyn}, \lambda=-1}(t) = -z \cos(\omega t)$
provides the explicit verification of the gauge bridge discussed above. Comparing this result
with the G\"oppert-Mayer transformation matrix defined in Eq.~\eqref{GM}, one finds that the
accumulated Peierls phase along either path corresponds exactly to the boundary phase mismatch
generated by the unitary operator $V^\dagger$ between the injection and detection ports. This
exact mathematical equivalence confirms that the dynamic Floquet-dressing phase can be seamlessly
translated into the language of discrete Peierls shifts at the junction nodes, maintaining strict gauge
invariance across the entire transport network.

Furthermore, because the linear-in-field dynamic phases acquired along the upper and lower arms are
completely identical, their relative contribution cancels out identically upon evaluating the interference
pattern at the output lead. This rigorous symmetry dictates
that the high-frequency modulation of the
current and the resulting contactless shutter functionality are driven exclusively by the quadratic-in-field
ponderomotive Magnus potential $K \propto E_0^2$ rather than linear phase fluctuations.

In realistic nanostructures, the quantum ring is electrostatically defined by external gates, while the injector
and detector ports are formed by quantum point contacts (QPCs) operating in the tunnel-barrier regime. In the
weak-coupling limit (characterized by high and wide potential barriers), an electron tunnels from the reservoir
lead into the ring with a small transmission amplitude,
$\mathcal{A}_{\text{in}}$. Under the asymmetric contact
configuration (positioned at a relative angle of $\pi/2$), the geometric paths are inherently
unequal: $L_{+1} = 3\pi R/2$ and $L_{-1} = \pi R/2$. In the purely ballistic transport regime without
dissipation, the transmission amplitude is independent of the path length, allowing us to safely assume
that the amplitudes along both arms are symmetric, $\mathcal{A}_{+1} = \mathcal{A}_{-1} = \mathcal{A}_{\text{in}}$.

Taking these boundary features into account, we define the partial wave function for each discrete
Floquet channel $n$ as follows:
\begin{equation}
\label{pfaz}
\Psi_s^{(n)} = \frac{1}{\sqrt{2}} \exp[-iz\cos(\omega t)] \cdot F_s^{(n)} \mathcal{A}_{\text{in}} \chi_s,
\end{equation}
where the qualitative quantum interference factor is explicitly given by:
\begin{equation}
\label{infaz}
F_s^{(n)} = \exp\left(i\frac{3\pi (C_n - sA)}{4}\right) - \exp\left(i\frac{\pi (C_n + sA)}{4}\right).
\end{equation}
The transmission probability within the $n$-th Floquet sideband is directly proportional to the squared
modulus of this interference factor:
\begin{equation}
\label{cfaz}
|F_s^{(n)}|^2 = 4\sin^2\left(\frac{\Delta \Phi_s^{(n)}}{2}\right), \quad \Delta \Phi_s^{(n)} = \pi\left(\frac{C_n}{2} - sA\right).
\end{equation}
A rigorous and complete description of the field-dressed states requires evaluating the dynamic phase modulation
present in Eq.~\eqref{pfaz} via the canonical Jacobi-Anger expansion:
\begin{equation}
\label{YaAn}
\exp[-iz\cos(\omega t)] = \sum_{n=-\infty}^{\infty} i^n J_n(z) e^{-in\omega t},
\end{equation}
where $J_n(z)$ is the $n$-th order Bessel function of the first kind. This algebraic expansion provides a
transparent physical insight into how the continuous dynamic phase drive breaks the continuous time-translation
symmetry, systematically distributing the single-particle probability density into an infinite manifold of
discrete, photon-assisted Floquet sidebands with effective quantum weights scaled by $J_n^2(z)$.

\section{Multichannel Transport in the Goppert-Mayer gauge: S-matrix approach}
\label{sec:transport}

\subsection{Physical Realizability and 1D Approximation}

To establish a solid bridge between the theoretical framework developed in Sec.~\ref{hamiltonian} and realistic experimental conditions, we evaluate the physical validity of the one-dimensional (1D) ring approximation using a concrete set of parameters optimized for high-mobility semiconductor heterostructures. Specifically, we consider a ternary $\text{In}_{0.53}\text{Ga}_{0.47}\text{As}$ quantum well channel, which is lattice-matched to an $\text{InP}$ substrate and known for hosting strong Rashba spin-orbit coupling. In our numerical calculations, the mean radius of the ring is chosen as $R = 200$~nm, the carrier effective mass is set to $m^* = 0.045\,m_e$ (where $m_e$ is the free electron mass), and the operating cryogenic temperature is maintained at $T = 100$~mK ($k_B T \approx 0.0086$~meV). The system is driven by a terahertz laser field with a photon energy of $\hbar\omega = 6$~meV, with the Fermi energy fixed at $E_F = 1.14$~meV. Under these specific operational conditions, the reduction of the finite-width physical channel to an effective 1D continuum is rigorously justified by the following criteria:

\begin{enumerate}
\item \textbf{Subband Frozen Dynamics (Transverse Quantization Boundary):}
For a realistic lithographic channel width of $W \sim 20$~nm, the lateral quantum confinement introduces a large energy splitting between the transverse size-quantized subbands, given by
$\Delta E_{\text{rad}} \approx \pi^2 \hbar^2 / (2m^* W^2) \approx 20.89$~meV. It is crucial to distinguish this scale from the longitudinal rotational quantum energy along the ring,
$\hbar\omega_0 = \hbar^2 / (2m^* R^2) \approx 0.02$~meV. Since $\Delta E_{\text{rad}} / \hbar\omega_0 = \pi^2 (R/W)^2 \gg 1$,
the transverse degree of freedom is completely decoupled from the azimuthal motion. The massive structural excitation energy $\Delta E_{\text{rad}}$ vastly exceeds the thermal scale ($k_B T \approx 0.0086$~meV), the Fermi energy
($E_F = 1.14$~meV $\approx 57\,\hbar\omega_0$),
and the driving terahertz photon energy
($\hbar\omega = 6$~meV $\approx 300\,\hbar\omega_0$).
Crucially, even for the highest active inelastic transport pathway considered in our model—the $n=2$ sideband, where the total carrier energy reaches $E_F + 2\hbar\omega \approx 13.14$~meV—the electronic energy remains safely below the transverse subband excitation threshold ($E_F + 2\hbar\omega < \Delta E_{\text{rad}}$). Consequently, all relevant elastic and multi-photon transport channels are strictly confined to the transverse ground-state subband, and inter-subband transition leakage is fundamentally blocked.

\item \textbf{Averaging of the Radial Rashba Component:}
    The strong spin-orbit interaction is characterized by the Rashba constant $\alpha_R = 14.195$~meV$\cdot$nm (see details in Sec.~\ref{details}). In a generic 2D channel, the Rashba Hamiltonian depends on both the azimuthal ($\hat{p}_\varphi$) and radial ($\hat{p}_r$) momentum operators. However, because the transverse wave function is rigidly confined within the ground-state subband of the narrow channel, the expectation value of the radial momentum vanishes identically ($\langle \hat{p}_r \rangle = 0$). The corresponding fluctuation term, $\langle \hat{p}_r^2 \rangle$, merely induces a static, uniform shift in the zero-point energy spectrum without breaking the spin symmetry. Thus, the spin-orbit dynamics are governed solely by the longitudinal momentum
    $\hat{p}_\varphi$, collapsing the multi-dimensional spin-orbit interaction into the exact 1D canonical form used in our transport equations.

    \item \textbf{Ballistic and Phase-Coherent Regime:}
    \textcolor{black}{In high-mobility InGaAs mesoscopic ring heterostructures at sub-Kelvin temperatures, the elastic mean free path has been measured to span $l_e \approx 1.4\text{--}1.7\,\mu\text{m}$ \cite{41} up to $l_e \approx 6\,\mu\text{m}$ \cite{42}, while the phase coherence length routinely reaches $l_\varphi \approx 4\text{--}6\,\mu\text{m}$ \cite{42}. Both scales substantially exceed the individual transit lengths along the ring arms ($L_{\text{down}} \approx 0.31\,\mu\text{m}$, $L_{\text{up}} \approx 0.94\,\mu\text{m}$) as well as the total circumference ($2\pi R \approx 1.25\,\mu\text{m}$). Consequently, electron transport is strictly ballistic and phase-coherent ($L_{\text{arm}} < 2\pi R < l_e \le l_\varphi$). Because the device operates far outside the diffusive scattering regime ($2\pi R < l_e$), disorder-induced phenomena such as universal conductance fluctuations (UCF) and Anderson localization are fundamentally suppressed. Furthermore, since higher transverse modes are energetically frozen out
    $(\Delta E_{\text{rad}}\gg E_F)$, minor elastic edge roughness does not trigger inter-subband mixing or phase randomization, and is
    fully absorbed into the effective transmission $(\mathcal{D}=\epsilon^2)$ and reflection $(r,t)$ parameters of the QPCs.}

\end{enumerate}

\noindent
Consequently, for the chosen parameters, all transverse degrees of freedom are fully frozen out. The idealized 1D ring model provides a physically sound, self-consistent, and experimentally realistic description of the quantum transport and spin-orbit interference phenomena under high-frequency THz driving, setting the stage for the formal multi-channel $S$-matrix construction detailed below.

\subsection{Advantages of the G\"oppert-Mayer Representation}

The application of the G\"oppert-Mayer (GM) unitary transformation~\cite{tan1}, which maps the system into the length gauge, constitutes a pivotal element of our theoretical framework. In contrast to conventional velocity-gauge formulations that require a formal renormalization of the carrier effective mass, the effective mass $m^*$ remains rigorously invariant within the length-gauge representation. Within the scattering region, the interaction with the driving electromagnetic field is absorbed into the time-periodic, spatially dependent phase factor of the wave function, which at the junction boundaries reduces to the discrete phase factor $\exp[-i z \cos(\omega t)]$.

Within the dipole approximation, which is strictly valid for the sub-wavelength dimensions of the nano-ring, the driving electric field is spatially homogeneous. Consequently, within the microscopic scattering zone of each QPC, the local gauge phase factor becomes spatially uniform. This local uniformity leads to a fundamental simplification of the transport problem: the quantum scattering processes occurring at the immediate vicinity of the QPCs can be described via a stationary, real-valued $S$-matrix. Thus, the complex high-frequency dynamics of the field-dressed electron at the junctions becomes mathematically equivalent to elastic scattering off a static potential barrier, rendering the local theoretical description highly transparent and self-consistent.

However, to establish a fully rigorous foundation for the quantum transport framework within the open network, it is imperative to elucidate how this length-gauge formulation circumvents the non-physical spatial divergences conventionally associated with infinite reservoirs. In a generic macroscopic system, the electric dipole potential, $V_{\text{dipole}} \propto \mathbf{r}\cdot\mathbf{E}$, exhibits a well-known pathological divergence due to its linear growth at infinity ($\mathbf{r} \to \infty$), which formally induces divergent energy shifts in the band structure of semi-infinite leads. In the proposed mesoscopic architecture, this non-physical divergence is resolved by two cooperative physical mechanisms:
\begin{enumerate}
\item \textbf{Topological Boundedness:} Within the ring itself, the continuous spatial coordinate is mapped onto the periodic azimuthal angle $\varphi$. Consequently, the interaction potential transforms into a strictly bounded trigonometric profile, $\mathcal{H}_{\text{int}}(t) = -V_0 \cos\varphi \sin(\omega t)$, which explicitly preserves the spatial periodicity and single-valuedness of the field-dressed wave functions, $\psi(\varphi) = \psi(\varphi + 2\pi)$, rendering the internal dynamics structurally protected from spatial divergences.
\item \textbf{Asymptotic Electrostatic Screening:} The subwavelength metallic leads and gating electrodes, acting as an effective plasmonic slot antenna, confine the high-frequency terahertz driving field to the central hotspot (see detailed discussion in Sec.~\ref{details}). Beyond the nano-ring region, the electromagnetic field undergoes rapid quasistatic depolarization screening, ensuring that the driving electric field asymptotically vanishes deep inside the reservoir channels ($\mathbf{E}_{\text{leads}} \to 0$).
\end{enumerate}

Mathematically, this spatial restriction establishes a robust ``gauge bridge'' at the injector ($\varphi=0$) and detector ($\varphi=3\pi/2$) ports. At these specific junction nodes, the unitary transformation operator $V^{\dagger}(t, \varphi)$ seamlessly matches the continuously dressed intra-ring states with the unperturbed Bloch states of the macroscopic leads. The accumulated line integrals of the vector potential across the junction boundaries reduce precisely to discrete, time-periodic phase factors conforming identically to the canonical Peierls substitution framework:
\begin{align}
  t_{d,k}(t)& = t_{d,k}^{(0)} \exp\left( \pm i \frac{e}{\hbar} \int_{\text{lead}}^{\text{ring}} \mathbf{A}(t) \cdot d\mathbf{l} \right)\nonumber\\
  & \equiv t_{d,k}^{(0)} \exp\left[ -i z \cos(\omega t) \right].
\end{align}
By decoupling the linear dipole drive from the asymptotic boundaries of the transport network, the high-frequency light-matter interaction is rigorously restricted to the active scattering region. This spatial separation ensures that the use of a stationary, real-valued $S$-matrix at the junctions remains strictly valid, thereby guaranteeing the strict gauge invariance, current continuity, and self-consistency of the Landauer--B\"uttiker multichannel transmission framework.

\subsection{Scattering Matrix and $\pi/2$ Geometry}
\label{Sgeom}

\begin{figure}[htp]
     \centering
     \includegraphics[width=0.35\textwidth,clip=]{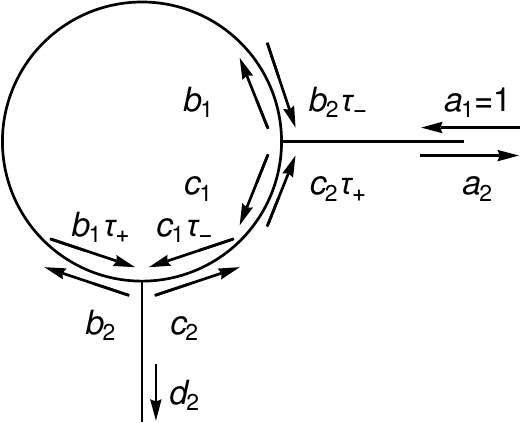}
     \caption{Definitions of electronic wave amplitudes of the quantum ring connected to
     two one-dimensional quantum transport leads.}
     \label{fig2}
\end{figure}

Let us analyze the quantum coupling junctions connecting the ring to the external reservoir leads (see Fig.~\ref{fig2}). We denote the internal electronic wave amplitudes within the ring channels as $b_j$ and $c_j$ ($j=1,2$). The parameters $r$ and $t$ represent the intra-ring reflection and transmission amplitudes of the QPCs, respectively, while $\sigma$ denotes the reflection amplitude from the reservoir lead back into itself. Finally, $\epsilon$ defines the coupling amplitude for a carrier transitioning between the lead and the ring.

Due to the time-reversal symmetry ($T$-invariance) of the system, the scattering amplitudes at the local Y-shaped junctions connecting the ring to the reservoir leads can be chosen as real-valued quantities. The symmetric $3\times3$ scattering matrix for an isolated three-terminal junction takes the form~\cite{but}:
\begin{equation}
S = \begin{pmatrix} r & \epsilon & t \\ \epsilon & \sigma & \epsilon \\ t & \epsilon & r \end{pmatrix},
\label{Smat}
\end{equation}
where the scattering parameters must satisfy the unitarity conditions of the $S$-matrix, ensuring the conservation of the total quantum current:
\begin{align}
r^2 + t^2 + \epsilon^2 &= 1, \label{unit1} \\
2\epsilon^2 + \sigma^2 &= 1. \label{unit2}
\end{align}

From these unitarity conditions, we derive the explicit analytical solutions for the scattering coefficients $r$, $t$, and $\sigma$:
\begin{align}
r &= \frac{1}{2} \left( \lambda_1 + \lambda_2 \sqrt{1-2 \epsilon^2} \right), \label{slrv} \\
t &= \frac{1}{2} \left( -\lambda_1 + \lambda_2 \sqrt{1-2 \epsilon^2} \right), \label{sltn} \\
&(r-t)^2 = 1, \label{rt1} \\
\sigma &= - \lambda_2 \sqrt{1 - 2 \epsilon^2}, \label{slsiv}
\end{align}
where $\lambda_1 = \pm 1$ and $\lambda_2 = \pm 1$ correspond to distinct algebraic branches dictating the boundary conditions. Under the physical requirement $\epsilon \le 1/\sqrt{2}$, these real-valued amplitudes directly imply two major algebraic invariants that hold regardless of the chosen sign branch:
\begin{equation}
r^2 + t^2 = 1 - \epsilon^2, \quad \text{and} \quad (r^2 - t^2)^2 = 1 - 2\epsilon^2.
\label{inv}
\end{equation}
The coupling strength (or openness) of the ring is determined by the single transmission parameter $\epsilon$, which lies within the bounded domain $\epsilon \in [0, 1/\sqrt{2}]$ to ensure that the reflection coefficient $\sigma$ remains strictly real. Finally, the total reflected and transmitted wave amplitudes in the external reservoir leads are denoted by $a_2$ and $d_2$, respectively.

The defining feature of the considered geometry (see Fig.~\ref{fig1}) lies in the structural asymmetry of the conducting arms. The phase accumulation factors acquired by the partial wave functions during their ballistic transit along the upper ($L_{\text{up}} = 3\pi R/2$) and lower ($L_{\text{down}} = \pi R/2$) arms for the $n$-th Floquet sideband and spin projection $s = \pm 1$ are explicitly defined as:
\begin{align}
\tau_1 &= \exp\left( i k_{+1, s}^{(n)} \cdot \Delta\varphi_1 \right) = \exp\left[ i 3\pi (C_n -  s A)/4 \right]\,, \nonumber \\
\tau_2 &= \exp\left( i k_{-1, s}^{(n)} \cdot \Delta\varphi_2 \right) = \exp\left[ i \pi ( C_n +  s A)/4 \right]. \label{t12}
\end{align}
From a physical standpoint, the sign inversion before the spin-orbit splitting terms in Eq.~\eqref{t12} (a minus sign for $\tau_1$ and a plus sign for $\tau_2$) directly captures the opposite sense of spatial circulation along the two conducting arms. The effective Rashba field $\mathbf{B}_{\mathrm{R}} \propto k_\varphi \mathbf{e}_r$ is oriented along the local radial direction in the plane of the ring, remaining strictly perpendicular to the electron's azimuthal momentum. As the carrier traverses the ring, this radial field vector continuously rotates along the perimeter, forming a closed in-plane spin texture. Crucially, the carrier traverses the upper long arm in the counterclockwise direction ($\Delta\varphi_1 > 0$), but propagates along the lower short arm in the clockwise direction ($\Delta\varphi_2 < 0$). Because the geometric Aharonov--Casher phase accumulated by a spin-$1/2$ particle depends strictly on the direction of contour integration through this spin texture ($\oint \mathbf{A}_{\text{SO}} \cdot d\mathbf{l}$), reversing the circulation path inverts the sign of the spin-orbit phase contribution, while the longitudinal kinetic phases remain positive along both paths. This directional sign asymmetry in the topological phase accumulation establishes the microscopic foundation for the spin phase mismatch at the detector.

  The effective field-dressed wave vectors $k_{\lambda,s}^{(n)}$ are taken directly from the analytical expression~\eqref{kls}. To determine the partial transmission amplitude $d_{2,s}^{(n)}$ and the system determinant $\mathcal{X}_s^{(n)}$, we employ the standard wave-function matching procedure at the junction nodes $\varphi=0$ and $\varphi=-\pi/2$, utilizing the local unitary $3\times3$ scattering matrix of the contacts defined in Eq.~\eqref{Smat}. Solving the resulting system of linear amplitude-balance equations via Cramer's rule, and systematically incorporating the unitary invariants~\eqref{inv}, we find that the partial amplitude describing the direct single-transit propagation takes the form:
\begin{align}
\label{d2s}
d_{2,s}^{(n)} = \frac{2\epsilon^2}{\mathcal{X}^{(n)}} \mathcal{N}_s^{(n)}\,.
\end{align}
Substituting the explicit expressions for the quantum point contact $S$-matrix elements and accounting for the structural asymmetry of the conducting arms ($L_{\text{up}}=3\pi R/2$, $L_{\text{down}}=\pi R/2$), the system determinant can be expressed in the following invariant trigonometric form:
\begin{align}
    \mathcal{X}^{(n)} = 1 - 2 \exp(i\pi C_n) \Big[ r^2 \cos\left(\frac{\pi}{2} C_n\right) \nonumber \\
    + t^2 \cos(\pi A) \Big] + (1 - 2\epsilon^2) \exp(i 2\pi C_n)\,.
 \label{Xs}
\end{align}
Equation~\eqref{Xs} demonstrates that the resonance spectrum, determined by the system determinant, is completely independent of the spin projection index $s$. Concurrently, the numerator is given by:
\begin{align}
\mathcal{N}_s^{(n)} = \cos\left(\frac{\pi}{4}C_n - \frac{\pi}{2}sA\right) - e^{i\pi C_n} \cos\left(\frac{\pi}{4}C_n + \frac{\pi}{2}sA\right)\,.
\label{eq:Nd_complex}
\end{align}
To evaluate the total transmission probability, we analyze the spin-dependent phase behavior within the numerator. Equation~\eqref{eq:Nd_complex} shows that the complex direct-transit amplitude $\mathcal{N}_s^{(n)}$ explicitly depends on the sign of the spin projection $s = \pm 1$ ($\uparrow, \downarrow$) via the topological Rashba phase parameter $A$. Consequently, although electrons with opposite spin projections simultaneously traverse the ring channels, their respective wave functions accumulate distinct complex phase shifts defined as $\vartheta_s = \arg(d_{2, s}^{(n)})$. This phase asymmetry reveals that the geometrically unbalanced arms of the ring act as a selective spin-phase rotator, mapping the topological phase onto spin-dependent phase arguments.

However, upon evaluating the absolute square of the scattering amplitude, which corresponds to the observable transmission probability, the algebraic structure of the numerator reduces to the following real-valued trigonometric expression:
\begin{align}
|\mathcal{N}_s^{(n)}|^2 &= 1 - \cos(\pi C_n) \cos(\pi A) \nonumber \\
&\quad + \cos \left(\frac{\pi}{2} C_n \right) \Big[ \cos(\pi A) - \cos(\pi C_n) \Big]\,. \label{eq:Nd_exact_final}
\end{align}

An analysis of Eq.~\eqref{eq:Nd_exact_final} reveals that the transmission probability $|\mathcal{N}_s^{(n)}|^2$ is strictly invariant with respect to the sign of the electron spin projection. Because the topological Rashba parameter $A$ enters Eq.~\eqref{eq:Nd_exact_final} exclusively within even trigonometric functions, the spin-flip operation ($s \to -s$) leaves the expression unchanged. This invariance establishes the strict spin degeneracy of the interferometer's transport envelope under unpolarized boundary conditions. Thus, combining the analytical expressions for both the numerator and the denominator yields the total transmission probability:
\begin{equation}
T_s^{(n)} = \frac{4\epsilon^4 |\mathcal{N}_s^{(n)}|^2}{|\mathcal{X}^{(n)}|^2} = T_n\,,
\label{tind}
\end{equation}
which formally confirms the vanishing net spin polarization of the total conductance.
\textcolor{black}{We note that higher-order coherent backscattering corrections, such as the $h/2e$-periodic Aronov--Altshuler--Spivak (AAS) effect \cite{AAS},
are strictly absent in our system. The AAS oscillations arise exclusively in the diffusive regime $(l_e\ll 2\pi R)$ due
to weak localization over time-reversed disordered loops. Because our clean single-mode InGaAs ring operates in the ballistic regime $(l_e > 2\pi R)$
single-transit interference dominates, while multi-round-trip trajectories are rigorously encapsulated within the Fabry--P\'erot denominator
$\mathcal{X}^{(n)}$ and filtered by the asymmetric Vernier geometry.}

\subsection{Factorization of the Direct Transmission Amplitude}
\label{factor}

To unveil the explicit analytical structure of the transmission spectrum and identify the exact nodes corresponding to complete channel blocking, we perform a rigorous algebraic factorization of the derived expression~\eqref{eq:Nd_exact_final}. Utilizing the double-angle trigonometric identity $\cos(\pi C_n) = 2\cos^2\left(\frac{\pi}{2} C_n\right) - 1$, we systematically group the terms containing the topological phase $\cos(\pi A)$ in Eq.~\eqref{eq:Nd_exact_final} as follows:
\begin{align}
|\mathcal{N}_s^{(n)}|^2 &= 1 - \cos(\pi C_n) \cos(\pi A) \nonumber \\
&\quad + \cos \left(\frac{\pi}{2} C_n \right) \Big[ \cos(\pi A) - \cos(\pi C_n) \Big] \nonumber \\
&= \left[ 1 + \cos\left(\frac{\pi}{2}C_n\right) - 2\cos^3\left(\frac{\pi}{2}C_n\right) \right] \nonumber \\
&\quad + \cos(\pi A) \left[ 1 + \cos\left(\frac{\pi}{2}C_n\right) - 2\cos^2\left(\frac{\pi}{2}C_n\right) \right]\,. \label{eq:factor_grouped}
\end{align}

By introducing the change of variable $x = \cos\left(\pi C_n/{2} \right)$, we observe that both polynomial blocks admit a rigorous algebraic factorization:
\begin{align}
1 + x - 2x^3 &= (1 - x)(2x^2 + 2x + 1), \\
1 + x - 2x^2 &= (1 - x)(2x + 1).
\end{align}
Factoring out the common element $(1-x)$ and returning to the original physical variables via the double-angle trigonometric identity $1 - \cos\left(\pi C_n/{2}\right) = 2\sin^2\left(\pi C_n/{4}\right)$, we arrive at the exact factorized form of the transmission numerator:
\begin{equation}
  |\mathcal{N}_s^{(n)}|^2 = 2\sin^2\left(\pi C_n/{4}\right) \cdot \mathcal{W}(x)\,,
\label{eq:Nd_factorized}
\end{equation}
where the structural factor $\mathcal{W}(x)$ is defined as a quadratic polynomial:
\begin{equation}
    \mathcal{W}(x) = 2x^2 + 2\Big[1 + \cos(\pi A)\Big]x + \Big[1 + \cos(\pi A)\Big]\,.
\label{eq:W_factor}
\end{equation}
The explicitly isolated first factor, $2\sin^2\left(\pi C_n/{4}\right)$, represents the classical quantum interference term for a direct electron transit along two distinct paths with a physical path-length difference of $\Delta L = \pi R$, which directly mirrors the foundational geometry of a Mach-Zehnder interferometer.

It is crucial to address a subtle quantum paradox inherent to this architecture: while the absolute squares of the channel transmissions appear to be independent of the spin projections, the entire functionality of the spin-protected optical rotator is determined uniquely by the phase arguments residing in the numerator of the scattering amplitude. In our model, the topological phase shift coexists with the time-periodic Floquet phase. It is the coherent quantum interference of these phase factors in the numerator—rather than a simple redistribution of transmission weights—that dictates the constructive or destructive nature of the wave-function superposition. When the THz field tunes the system toward the resonance condition, these complex phases in the numerator undergo sharp synchronization, giving rise to ultra-narrow, delta-like Airy spikes that remain highly resilient to ambient thermal fluctuations.

This mechanism provides a distinct technological advantage over the Datta–Das spin transistor~\cite{datta}. In the classical Datta–Das geometry, spin precession is controlled via a static electric field utilizing the Rashba spin-orbit interaction. Consequently, the device suffers from thermal velocity dispersion and phase smearing, because electrons within the thermal Fermi distribution possess different wave vectors and experience varying precession rates, degrading the switching characteristics at finite temperatures. In stark contrast, our proposed THz quantum shutter overcomes this fundamental limitation. Since the switching functionality is dynamically locked by the ponderomotive Floquet parameter via the $J_0(z) = 0$ collapse, the critical threshold is anchored strictly to the coherent driving frequency of the external THz field rather than the stochastic thermal energy of the carriers. This dynamic phase-locking ensures that the sharp switching threshold is preserved even at finite temperatures, representing a major conceptual advancement over conventional static spin-orbit devices.
%%%%%%%%     Phase  %%%%%%%%%%%%

\subsection{Topological Robustness of the Macroscopic Envelope}
\label{topstab}

An analysis of the structural factor $\mathcal{W}(x)$ reveals a fundamental property of the proposed interferometer. Let us evaluate the discriminant of this quadratic polynomial with respect to the variable $x \in [-1, 1]$:
\begin{equation}
    \Delta = 4\Big(1 + \cos(\pi A)\Big)^2 - 8\Big(1 + \cos(\pi A)\Big) = 4\Big(\cos^2(\pi A) - 1\Big).
\label{eq:discriminant}
\end{equation}
Given that for any real values of the spin-orbit interaction parameter the constraint $\cos^2(\pi A) \le 1$ holds universally, the discriminant is always non-positive ($\Delta \le 0$). It vanishes exclusively at highly degenerate points of symmetry, while remaining strictly negative under off-resonance parameter configurations. This mathematical feature guarantees that the polynomial $\mathcal{W}(x)$ is strictly positive-definite and devoid of zero-crossings within the physical domain of definition, ensuring stability across the entire interval of the variable.

This algebraic fact carries a critical physical consequence: the macroscopic transmission envelope exhibits robust topological stability. The quantum system is fundamentally immune to the emergence of any parasitic conductance zeroes upon arbitrary variations of the Rashba spin-orbit parameter $A$. Instead, the spin-orbit interaction acts exclusively as a smooth amplitude modulator of the global transport envelope.

In our model for an $\text{InGaAs}$ ring with a radius of $R = 200$~nm, the spin-orbit interaction parameter is fixed at $A = 3.5$ ($\alpha_R \approx 14.195$~meV$\cdot$nm), which aligns well with experimentally accessible values achieved via external gate voltage tuning~\cite{man}. For this specific operational point, $\cos(3.5\pi) = 0$, which simplifies the structural polynomial to $\mathcal{W}(x) = 2x^2 + 2x + 1$, while its discriminant becomes $\Delta = -4 < 0$.

The chosen configuration is optimal, as it guarantees the maximum dynamic range for transport control. All nodes corresponding to complete channel blocking remain locked strictly to both the geometric quantum interference condition, $\sin^2\left(\pi C_n/{4}\right) = 0$, and the external high-frequency Floquet gate condition, $J_0(z) = 0$.

 \subsection{Coherent Spin Dynamics and Implementation of the Datta--Das Paradigm}
 \label{DDas}

The spin-valve functionality of the proposed device is realized by implementing
the paradigm of the Datta--Das spin field-effect transistor~\cite{datta}. We
assume that the injector and detector leads are ferromagnetic, with collinear
magnetizations oriented along the $x$-axis (lying in the plane of the quantum ring). This configuration is required because the effective spin-orbit Rashba magnetic field, generated by the structural inversion asymmetry, inherently lies within the plane of the ring and rotates continuously with the electron's azimuthal momentum due to the spin-momentum locking mechanism. Specifically, since the carrier is injected along the $x$-axis, the resulting effective Rashba field is oriented along the $y$-axis at the injection junction ($\varphi = 0$).

With the incoming carriers polarized along the $x$-direction, their initial
spins are strictly perpendicular to the local radial Rashba field at the entrance. To analyze the transport within the rotating coordinate frame, the spin-polarized electron injected from the ferromagnetic lead is described by the pure quantum state $\vert +x \rangle$, which is represented as a coherent superposition of the $\sigma_z$ eigenstates of the internal framework:
\begin{equation}
   \vert \Psi_{\text{in}} \rangle = \vert +x \rangle = \frac{1}{\sqrt{2}} \Big( \vert \uparrow_z \rangle + \vert \downarrow_z \rangle \Big).
\end{equation}

At this point, a methodological remark regarding the lead--ring interface is warranted. In a realistic hybrid device, the direct coupling between a ferromagnetic metal lead and an InGaAs semiconductor channel introduces a severe conductivity mismatch, characterized by a major Fermi wavevector difference ($k_{\rm F}^{\rm lead} \gg k_{\rm F}^{\rm ring}$)
alongside a spin-dependent density of states. This mismatch typically induces strong spin-selective backscattering, which suppresses the spin injection efficiency. To circumvent this conductivity mismatch problem and preserve the coherent, spin-degenerate transport dynamics within the loop, we assume that the contacts are implemented via high-resistance tunnel junctions
(e.g., thin $\text{Al}_2\text{O}_3$ barriers or engineered Schottky interfaces) operating in the low-transparency regime
($\mathcal{D} = \epsilon^2 \ll 1$).

Under this interfacial constraint, the high potential barrier dominates the
boundary-matching conditions. The internal reflection amplitude from the contact
interface becomes structurally clamped close to unity for both spin projections
($r_{\uparrow} \approx r_{\downarrow} \approx 1$), effectively rendering the
internal Fabry--P{\'e}rot multi-round-trip parameter $\mathcal{B} = 1 - 2\epsilon^2$
strictly spin-independent. Consequently, at the injection node ($\varphi = 0$),
the spatial spin-rotation operator for the transmitted wave successfully reduces
to the identity matrix, $U(0) = \hat{\mathbb{I}}$. This approach filters out
parasitic spin-dependent backscattering within the cavity, ensuring a coherent
transition of the spin wave function between the lead state and the quantum ring
channel, while the ferromagnetic nature of the contacts manifests purely through
the spin projection coefficients during injection and detection.

As the carrier propagates ballistically through the arms of the asymmetric
interferometer, the coherent spin precession---characterized by the expectation
value of the spin polarization vector---is strictly confined to the $xy$-plane
of the ring ($\langle S_z \rangle = 0$). Each spin component propagates
independently and acquires its respective complex transmission amplitude
$d_{2,s}^{(n)}$, where $s = \{\uparrow, \downarrow\}$ denotes the spin projection along the structural $z$-axis. Owing to the generalized time-reversal invariance governing the Floquet quantum evolution in the absence of external magnetic fields, the non-equilibrium driving does not lift the underlying dynamic spectral degeneracy between the spin components. Consequently, these generalized temporal symmetry constraints dictate that the absolute magnitudes of these partial transmission amplitudes must remain strictly equal across both spin projections:
\begin{equation}
   \vert d_{2,\uparrow}^{(n)} \vert = \vert d_{2,\downarrow}^{(n)} \vert = \sqrt{T_n}\,.
\end{equation}
Here, $T_n$ represents the spin-degenerate macroscopic transport envelope of the
$n$-th Floquet channel. Since the external THz laser does not introduce any
time-dependent Zeeman interactions that could directly couple to the spin degree
of freedom, this dynamic phase modulation of the orbital states leaves the magnitude degeneracy unperturbed over the driving period.

However, their complex phases are distinct: the spin-dependent phase shift
$\vartheta_{s}^{(n)} = \arg(d_{2,s}^{(n)})$ accumulates because the spin-dependent Rashba term $sA$ explicitly enters the arguments of the trigonometric functions in the scattering numerator $\mathcal{N}_s^{(n)}$ (see Eq.~\eqref{eq:Nd_complex}).
Representing these partial amplitudes in the polar form $d_{2,s}^{(n)} = \sqrt{T_n} e^{i\vartheta_{s}^{(n)}}$,
the emergent partial spinor state corresponding to the $n$-th Floquet sideband,
evaluated immediately prior to the detecting ferromagnetic drain, is expressed as:
\begin{equation}
  \vert \Psi_{\text{out}}^{(n)} \rangle = \frac{\sqrt{T_n}}{\sqrt{2}} \Big( e^{i\vartheta_{\uparrow}^{(n)}} \vert \uparrow_z \rangle + e^{i\vartheta_{\downarrow}^{(n)}} \vert \downarrow_z \rangle \Big).
  \label{eq:psi_out_z}
\end{equation}
The phase mismatch
$\vartheta_{\uparrow}^{(n)} \neq \vartheta_{\downarrow}^{(n)}$
physically manifests as a coherent spin precession of the carrier polarization
vector within the $xy$-plane during its ballistic transit. To formulate the
transport characteristics under realistic experimental conditions, the detecting
ferromagnetic drain is characterized by a finite spin polarization efficiency
$P_{\rm lead} = (N_{\to} - N_{\gets}) / (N_{\to} + N_{\gets})$, where
$N_{\to(\gets)}$ represents the density of states for spins aligned parallel
(antiparallel) to the $x$-axis. The total spin-filtered transmission probability
within the $n$-th Floquet sideband is determined by projecting the emergent
partial spinor state $\vert\Psi_{\text{out}}^{(n)}\rangle$ onto the collinear
basis states $\vert \pm x \rangle = \frac{1}{\sqrt{2}} \big( \vert \uparrow_z \rangle \pm \vert \downarrow_z \rangle \big)$:
\begin{equation}
    T_{\text{FM}}^{(n)} = (1 + P_{\rm lead}) \left| \langle +x \vert \Psi_{\text{out}}^{(n)} \rangle \right|^2 + (1 - P_{\rm lead}) \left| \langle -x \vert \Psi_{\text{out}}^{(n)} \rangle \right|^2.
\end{equation}
By evaluating the spinor projections, the total transmission probability can be
expressed in terms of the absolute squares of the sums and differences of the
complex phase factors:
\begin{align}
    T_{\rm FM}^{(n)} &= \frac{T_n}{4} (1 + P_{\rm lead}) \left| e^{i\vartheta_{\uparrow}^{(n)}} + e^{i\vartheta_{\downarrow}^{(n)}} \right|^2  \nonumber\\
    &+ \frac{T_n}{4} (1 - P_{\rm lead}) \left| e^{i\vartheta_{\uparrow}^{(n)}} - e^{i\vartheta_{\downarrow}^{(n)}} \right|^2.
    \label{T_FM_intermediate}
\end{align}
By applying the Euler identities and half-angle trigonometric relations to the
algebraic structure of Eq.~\eqref{T_FM_intermediate}, the interference terms
expand into $\cos^2[(\vartheta_{\uparrow}^{(n)} - \vartheta_{\downarrow}^{(n)})/2]$
and $\sin^2[(\vartheta_{\uparrow}^{(n)} - \vartheta_{\downarrow}^{(n)})/2]$, respectively.
Utilizing the fundamental identity $\sin^2(x) = 1 - \cos^2(x)$, the expression
simplifies to its final analytical form:
\begin{align}
    T_{\rm FM}^{(n)} &= T_n \left[ \frac{1 - P_{\rm lead}}{2} + P_{\rm lead} \cos^2\left(\frac{\vartheta_{\uparrow}^{(n)} - \vartheta_{\downarrow}^{(n)}}{2}\right) \right] \nonumber\\
    &= T_n \left[ \frac{1 - P_{\rm lead}}{2} + P_{\rm lead} \cos^2\left(\frac{\Delta\vartheta_n}{2}\right) \right],
    \label{T_FM_final}
\end{align}
where $\Delta\vartheta_n \equiv \vartheta_{\uparrow}^{(n)} - \vartheta_{\downarrow}^{(n)}$
represents the intensity-dependent differential spin phase. In the idealized limit
of a half-metallic collector ($P_{\rm lead} \to 1$), Eq.~\eqref{T_FM_final} smoothly
converges to the strict Datta--Das modulation envelope, $T_{\rm FM}^{(n)} = T_n \cos^2(\Delta\vartheta_n / 2)$.
In the device analysis presented in Sec.~V, this finite polarization acts as a strict
visibility scaling factor, ensuring that the dynamic phase difference $\Delta\vartheta_n$
serves as the primary control parameter modulating the amplitude of the output
spin-polarized current.

This central result effectively factorizes the operational physics of the
optically driven device into two distinct, functionally orthogonal quantum
modulation mechanisms. First, an ``amplitude valve'' ($T_n$) governs the global,
spin-degenerate charge transparency of the open cavity, where the external THz
field is capable of driving the system into a completely closed state via the
$J_0(z_c) = 0$ dynamic localization threshold. Second, a ``spin shutter''
(acting as a spin-phase modulator) emerges, described by the modulated transmission
envelope established in Eq.~\eqref{T_FM_final}, wherein the static Rashba spin-orbit coupling acts in tandem with
the phase sensitivity near an Exceptional Point (EP; see detailed discussion in Sec.~\ref{subsec:discussion} and Appendix~B)
to predefine and dynamically rotate the coherent electron spin precession angle.
Although both mechanisms are nonlinearly coupled via their shared dependence on
the laser-induced ponderomotive shift $K(I)$, they control fundamentally different
degrees of freedom in the output current.

The presence of a non-zero, intensity-dependent phase difference $\Delta\vartheta_n$,
embedded within the complex algebraic structure of the scattering numerators,
enables the proposed architecture to operate as a fully functional, high-frequency Datta--Das spin modulator within a spin-valve junction. Importantly, this mechanism remains strictly consistent with fundamental symmetry prohibitions regarding the non-magnetic generation of spin-polarized currents. In accordance with the generalized Landauer--B{\"u}ttiker transport formalism, the Onsager reciprocal relations for a spin-degenerate system restrict asymmetric spin transmission within isolated non-magnetic cavities. Here, these time-reversal symmetry constraints are successfully circumvented by the magnetic boundary conditions of the measurement circuit itself. The ferromagnetic boundary conditions of the drain lift the macroscopic spin degeneracy via non-equilibrium quantum projection, converting the purely phase-encoded spin precession accumulated inside the ring into a measurable, highly modulated
spin-polarized charge current.

\section{Conductance and Multichannel Transport}
\label{provod}

According to the generalized Landauer--B\"uttiker formulation for periodically driven systems~\cite{butm,moscal}, the total time-averaged conductance for the ferromagnetic configuration is obtained by summing the partial contributions from all accessible photon channels (Floquet sidebands):
\begin{equation}
G_{\text{FM}}(z) = G_0 \sum_{n=-\infty}^{\infty} J_n^2(z) T_{\text{FM}}^{(n)}(E_F + n\hbar\omega),
\label{Gs}
\end{equation}
where $G_0 = e^2/h$ is the fundamental conductance quantum, and $T_{\text{FM}}^{(n)}(E_F + n\hbar\omega)$ represents the spin-filtered electron transmission probability within the $n$-th channel at the sideband-shifted energy, defined by Eq.~\eqref{T_FM_final}.

While Eq.~\eqref{Gs} is formally written in the zero-temperature limit, its application remains valid and highly accurate for realistic sub-Kelvin operational conditions ($T \le 100$~mK). At these experimental temperatures, the characteristic thermal energy scale is negligibly small, $k_{\text{B}} T \approx 0.0086$~meV. This thermal scale is heavily suppressed compared to the internal spatial quantization energy of the quantum ring, $\hbar\omega_0 \approx 0.02$~meV, and is nearly three orders of magnitude smaller than the driving photon energy, $\hbar\omega = 6$~meV. Under this pronounced energy hierarchy ($k_{\text{B}} T \ll \hbar\omega_0 \ll \hbar\omega$), the derivative of the Fermi--Dirac distribution function reduces to a sharp Dirac delta function, $-\partial f_0/\partial E \to \delta(E - E_{\text{F}})$, thereby eliminating any parasitic thermal smearing of the quantum interference fringes and justifying the formulation~\eqref{Gs}.

Meanwhile, it is possible to truncate the infinite Floquet sum , which physically is justified by two main circumstances.
First, within the considered THz frequency range
($\hbar\omega = 6$~meV) and at the selected electron Fermi energy $E_{\rm F} \approx 1.14$~meV, all transport processes involving the emission of driving photons ($n < 0$) are fully suppressed. The available kinetic energy for these negative-index channels becomes negative ($E_{\rm F} - \lvert n \rvert\hbar\omega < 0$), which renders their corresponding wave vectors strictly imaginary. Consequently, these channels are converted into evanescent modes that decay exponentially away from the contact interfaces. This energetic cutoff rigorously allows us to restrict the remaining Floquet series to non-negative harmonics ($n \ge 0$).
Consequently, the total time-averaged multi-channel transport response is governed by the primary elastic signal ($n=0$) accompanied by coherent inelastic background contributions originating from the lowest-order photon-assisted sidebands ($n=1, 2$):
\begin{align}
    G_{\rm FM}(I) &\approx G_0 \Big[ J_0^2(z) T_{\rm FM}^{(0)}(E_{\rm F}) + J_1^2(z) T_{\rm FM}^{(1)}(E_{\rm F}+\hbar\omega) \nonumber\\
    &+ J_2^2(z) T_{\rm FM}^{(2)}(E_{\rm F}+2\hbar\omega) \Big].
    \label{eq:G_FM_channels_sum}
\end{align}
Eq.~\eqref{eq:G_FM_channels_sum} serves as the foundation for evaluating the spin-resolved parallel ($G_{\rm P}$) and antiparallel ($G_{\rm AP}$) conductance profiles analyzed under explicit ferromagnetic boundary conditions in the subsequent sections of this study.
\textcolor{black}{Quantitatively, the neglect of higher-order sidebands ($n\geq 3$)
 is rigorously justified by the large energy hierarchy $\hbar\omega/\Gamma\approx 405$.  Even under a conservative Lorentzian cavity-tail estimate,
the off-resonance transmission probability for the $n=3$ channel is suppressed by $(\Gamma/(6\hbar\omega))^2 \approx 1.7\times 10^{-7}$.
Weighted by the Bessel factor $J_3^2(z_c)\approx 0.038$ at the critical threshold, its contribution to the conductance is $G_3 < 10^{-8}G_0$,
which is four orders of magnitude smaller than the baseline noise floor $\sigma_{\text{bg}}^2=5\times 10^{-4}$
(see detailed discussion in Sec.~\ref{subsec:snr}). Consequently, truncating the multichannel expansion
at $n=2$ is an exceptionally accurate approximation across the entire operational intensity range.}

Second, this truncation of the Floquet hierarchy is supported by a
well-defined hierarchy of energy scales. The driving frequency of $1.45$~THz
($\hbar\omega = 6$~meV) provides a sufficiently large spectral separation to
ensure strong isolation between adjacent Floquet sidebands. A straightforward
estimation of the intrinsic resonance linewidth can be obtained from the contact
transparency parameter, defined as $\mathcal{D} = \epsilon^2$. This coupling
parameter $\mathcal{D}$ determines the effective width of the Fabry--P{\'e}rot
resonant peaks as $\Gamma = 2\mathcal{D} E_{\rm Th}$, where the Thouless energy,
\begin{equation}
E_{\rm Th} = \frac{\hbar}{2\pi R}\sqrt{\frac{2E_{\rm F}}{m^*}} = \frac{\hbar}{\tau}_{\text{orb}}\,.
\label{E_Th}
\end{equation}
It serves as the characteristic energy scale associated with the ballistic
round-trip transit time $\tau_{\text{orb}} = 2\pi R / v_{\rm F} \approx 13.3$ps of an electron completing a
closed orbit around the loop. For our concrete weak-coupling scenario with a
tunnel transparency of $\mathcal{D} = 0.15$ and an operating Fermi energy of
$E_{\rm F} = 1.14$~meV, the parameters self-consistently yield $E_{\rm Th} \approx 0.0494$~meV. The resulting intrinsic resonance width is thus evaluated as $\Gamma \approx 0.0148$~meV.

This finite width implies that the trapped electron undergoes an average of
$\tau_{\rm life} / \tau = 1 / (2\mathcal{D}) \approx 3.33$ complete ballistic
round-trips before escaping into the reservoirs, which is sufficient to establish multi-wave quantum interference. Crucially, at a cryogenic base configuration of $T = 100$~mK, the background thermal energy scale is frozen at $k_{\rm B} T \approx 0.0086$~meV, remaining strictly below the intrinsic linewidth ($\Gamma \approx 1.72 \, k_{\rm B} T$). Although continuous THz driving induces steady-state intra-band free-carrier absorption,
the ultra-low operating Fermi energy severely restricts the phase space available for electronic photo-excitation. Furthermore, the coupling to the semi-infinite leads ensures that the massive reservoirs act as ideal, non-equilibrium heat sinks, rapidly evacuating hot carriers. This efficient thermal relaxation prevents the non-equilibrium electronic temperature $T_{\rm e}$ from exceeding the coherence threshold ($k_{\rm B} T_{\rm e} < \Gamma$), guaranteeing that the sharp Fabry--P{\'e}rot resonance peaks remain resilient against thermal smearing during the active modulation cycle.

Since the photon energy of the high-frequency field ($\hbar\omega = 6$~meV) exceeds the intrinsic width of the quantum ring's interference resonances
($\Gamma \approx 0.0148$~meV) by more than two orders of magnitude, the system satisfies the fundamental inequality $\hbar\omega \gg \Gamma$. Within this off-resonance high-frequency approximation, the resonant Fabry--P{\'e}rot manifolds belonging to adjacent Floquet sidebands become spectrally isolated from one another. Consequently, any cross-channel quantum coherence between electrons modulated into different sidebands is heavily suppressed upon energy-integration over the transport window. This separation ensures that while the total multichannel scattering matrix retains its intrinsic
inelastic coupling, the time-averaged conductance (and thus the total current)
effectively decouples from phase-dependent cross-channel interference.
Under these asymptotically decoupled transport conditions, the total charge and
spin transport reduces to an independent sum over the individual Floquet channel
transmission probabilities, $T_{\rm FM}^{(n)}$. Within each isolated sideband,
the driving field manifests as an effective source modulator, where the
transmission profile is rigorously weighted by the squared Bessel functions
$J_n^2(z)$ that emerge from the Jacobi--Anger expansion of the Floquet states.
This structural factorization isolates the energetic pathways within individual
photon sidebands, rendering the theoretical description of the driven mesoscopic
network highly robust, free from spurious inter-channel interference, and
conceptually complete.

\subsection{Hybrid Interference Architecture: Mach--Zehnder and Coupled Fabry--P\'erot Cavities}
\label{sec:Hybrid_Interference}

The spin-degenerate transmission probability in the $n$-th Floquet channel,
derived using the $S$-matrix formalism, is expressed as $T_n = 4\epsilon^4 \lvert \mathcal{N}_s^{(n)} \rvert^2 / \lvert \mathcal{X}^{(n)} \rvert^2$.
\textcolor{black}{To understand the operational physics of the asymmetric ring, it is instructive to explicitly delineate the direct electronic analogue of the optical Vernier effect from the spin-orbit scattering mechanisms:}\\
\textcolor{black}{
(i) {\bf Electronic Vernier Filtering (Orbital Cavity Dynamics):} In close analogy to cascaded optical microring filters, the asymmetric
3:1 path-length configuration $(L_{\text{up}}=3\pi R/2$, $L_{\text{down}}=\pi R/2)$
establishes two coupled Fabry--P\'erot sub-cavities with mismatched free spectral ranges (FSR$_{\text{down}}$=3FSR$_{\text{up}}$).
 As demonstrated by the algebraic factorization of the system determinant
 $\lvert \mathcal{X}^{(n)} \rvert^2=\lvert Z_1 \rvert^2 \cdot \lvert Z_2 \rvert^2$ (see Appendix~A),
 this spatial asymmetry suppresses off-resonance background continuum modes while selectively sustaining ultra-sharp Airy resonances with a high effective finesse $(\mathcal{F}\approx 125)$. Importantly, because the system determinant is strictly spin-degenerate, this Vernier effect acts purely as a spatial and spectral orbital charge filter.}\\
\textcolor{black}{
(ii) {\bf Rashba Spin-Orbit Modulation (Spin-Phase Rotation):} The spin-dependent functionality is mediated exclusively by the scattering numerator
 $\mathcal{N}_{s}^{(n)}$ $[Eq.(43)]$.  Due to the relativistic spin-momentum locking, the geometric Rashba parameter $sA$ enters the arguments of the counter-propagating partial waves with opposite signs $\mp sA\pi/2$. This directional phase splitting maps the Vernier-sharpened orbital transmission grid onto a coherent spin phase rotator, enabling high-contrast Datta--Das modulation at the ferromagnetic collector.}

Here, the system determinant $\mathcal{X}^{(n)}$ is strictly spin-degenerate,
while the spin dependence of the scattering numerator $\mathcal{N}_s^{(n)}$
vanishes identically upon evaluating its absolute square due to time-reversal
symmetry, ensuring that $T_n$ serves as a macroscopic transport envelope.
This exact analytical solution allows for a detailed decomposition of the quantum interference mechanisms inside
the mesoscopic ring. In contrast to simplified phenomenological models, this formulation explicitly
demonstrates the hybrid nature of the device, where spatial path interference between the two geometric arms
(akin to a Mach--Zehnder interferometer) continuously modulates the resonant excitation of the multi-round-trip internal modes (characteristic of coupled Fabry--P{\'e}rot sub-cavities).

\subsubsection{The Numerator: The Mach--Zehnder Envelope}

The transmission numerator $\mathcal{N}_s^{(n)}$ captures the quantum interference of wave amplitudes propagating along the upper and lower arms of the ring. Within the scattering formalism, the squared modulus of this numerator, $\lvert \mathcal{N}_s^{(n)} \rvert^2$, predefines the smooth transmission envelope of the spectrum, conceptually analogous
to the response of a Mach--Zehnder interferometer. Crucially, while the individual complex amplitudes $\mathcal{N}_s^{(n)}$ possess spin-dependent phases due to the Rashba spin-momentum locking, their absolute squared moduli $\lvert \mathcal{N}_s^{(n)} \rvert^2$ remain strictly invariant under the spin-reversal operation ($s \to -s$), whereas the system determinant
 $\mathcal{X}^{(n)}$ is intrinsically spin-independent.

Consequently, owing to this structural symmetry, a two-terminal ring geometry
(even with structurally asymmetric arms) connected to unpolarized reservoirs
cannot generate a net spin-polarized current on its own. Instead, the topological Aharonov--Casher phase, entering via the $\cos(\pi A)$ terms, manifests purely as a global charge amplitude modulator of the broad Mach--Zehnder envelope. The latent spin-dependent phase information encoded within the unsquared scattering numerators is only read out and converted into a measurable spin polarization when the emergent spinor states undergo quantum projection at the ferromagnetic collector interface, as derived in Sec.~\ref{DDas}.

\subsubsection{The Denominator: Coupled Fabry--P\'erot Cavities Regime}

The electronic states and resonant properties of the open ring, governed by the
contact barriers of finite coupling amplitude $\epsilon$, are fully encapsulated
by the system determinant $\mathcal{X}^{(n)}$. As established via the exact
boundary-matching expression, the transmission denominator $\lvert\mathcal{X}^{(n)}\rvert^2$ is determined by the squared modulus of the determinant, which can be structurally formulated as an effective quadratic-like function with respect to the round-trip phase factor $y = \exp(i\pi C_n)$ (see Eq.~\eqref{x1} in Appendix~\ref{FP_factor}):
\begin{equation*}
\mathcal{X}^{(n)} = 1 - \mathcal{A}(C_n, A) y + \mathcal{B} y^2\,,
\end{equation*}
where the coefficient $\mathcal{A}(C_n, A) = 2 \left[ r^2 \cos\left(\pi C_n/2\right) + t^2 \cos(\pi A) \right]$
acts as the effective internal reflection coefficient, and $\mathcal{B} = 1 - 2\epsilon^2$ represents the loop-confinement parameter.

Crucially, as detailed in Appendix~\ref{FP_factor}, since the internal reflection
coefficient $\mathcal{A}(C_n, A)$ nonlinearly depends on the energy via the
half-round-trip phase contribution $\cos(\pi C_n/2)$, Eq.~\eqref{x1}
is not a simple quadratic polynomial. Instead, upon introducing the fundamental
half-round-trip phase variable $u = \exp(i\pi C_n / 2)$ (where $y = u^2$), the
resonance condition $\mathcal{X}^{(n)} = 0$ rigorously maps onto a reciprocal-like
polynomial of the fourth degree. This exact fourth-order algebraic structure
fully accounts for the multi-wave interference and the counter-propagating modes
confined within the coupled sub-cavities of the open ring interferometer.

Nevertheless, owing to the structural symmetry of this reciprocal fourth-order
system under the optimal operational constraints, the determinant can be
self-consistently factorized into a product of twin effective Fabry--P{\'e}rot
sub-cavity responses. By introducing the effective complex loop-reflection
parameters $\mathcal{R}_1$ and $\mathcal{R}_2$, the system determinant is
rigorously expressed in the factored form (see Eq.~\eqref{eq:Xs_fact_app} in
Appendix~\ref{FP_factor}):
\begin{equation*}
 \mathcal{X}^{(n)} = (1 - \mathcal{R}_1 y)(1 - \mathcal{R}_2 y) = \left(1 - \mathcal{R}_1 e^{i\pi C_n}\right)\left(1 - \mathcal{R}_2 e^{i\pi C_n}\right)\,,
\end{equation*}
where, as a direct consequence of the algebraic symmetry relations of the underlying
quartic polynomial, the product of these effective complex reflection parameters
remains uniquely clamped by the tunnel transparency of the contacts to the value
$\mathcal{R}_1 \mathcal{R}_2 = \mathcal{B} = 1 - 2\epsilon^2$.

The squared modulus of each linear factor in Eq.~\eqref{eq:Xs_fact_app} can be
rigorously mapped onto the canonical form of a standard Fabry--P{\'e}rot resonance
denominator up to a global scaling factor. By expressing the complex
loop-reflection parameters in polar form as $\mathcal{R}_j = \lvert\mathcal{R}_j\rvert e^{i\phi_j}$,
whose explicit dependence on the junction scattering parameters and energy dependence
is systematically derived in Appendix~\ref{FP_factor}, the algebraic expansion yields:
\begin{equation}
\left| 1 - \mathcal{R}_j e^{i\pi C_n} \right|^2 = (1 - \lvert\mathcal{R}_j\rvert)^2 \left[ 1 + \mathcal{F}_j \sin^2\left(\frac{\pi C_n}{2} + \delta_j\right) \right],
\label{eq:Airy_canonical}
\end{equation}
where the effective finesse parameter is rigorously defined as $\mathcal{F}_j = 4\lvert\mathcal{R}_j\rvert / (1 - \lvert\mathcal{R}_j\rvert)^2$,
and the internal resonant phase shift is clamped to exactly half of the complex
loop-reflection parameter argument, $\delta_j = \phi_j / 2$.

As shown in the algebraic analysis of Appendix~\ref{FP_factor}, within the
mutual lock-in regime, the moduli are strictly clamped to the loop-confinement parameter
($\lvert\mathcal{R}_1\rvert = \lvert\mathcal{R}_2\rvert = \sqrt{\mathcal{B}}$),
which ensures identical finesse values for both effective sub-cavities, while
their phases satisfy $\phi_1 = -\phi_2 = \theta$. Consequently, this microscopic
transport decomposition explicitly demonstrates that due to the structural
arm-length asymmetry ($L_{\rm up}/L_{\rm down} = 3$), the open quantum ring
functions not as a single isolated optical etalon, but rather as a non-trivial
system of two coupled Fabry--P{\'e}rot sub-cavities. The physical emergence of
this internal coupled-resonance regime is driven by the coherent backscattering
of ballistic electrons at the contact junctions into the shorter geometric arm
(subtending an azimuth angle of $\pi/2$). This localized reflection process
self-consistently generates the kinematic interference term $\cos(\pi C_n/2)$
within the effective internal reflection coefficient $\mathcal{A}$, lifting
the single-cavity paradigm.

\subsubsection{Multichannel Routing and the Spin Shutter Mechanism}

The hybrid interferometer architecture analyzed above---combining a Mach--Zehnder
configuration with coupled Fabry--P{\'e}rot sub-cavities---characterizes the
non-equilibrium quantum scattering dynamics within a single, isolated Floquet
channel. However, the total spin-resolved transport response of the hybrid device
is determined by the independent summation of decoupled contributions from all
energetically accessible transport pathways, as described by the multi-channel
Landauer--B{\"u}ttiker relation. Within the investigated cryogenic temperature
and low-energy regime ($E_{\rm F} = 1.14$~meV), the dominant transport channels
are strictly restricted to the central elastic channel ($n = 0$) and the
lowest-order inelastic sidebands ($n = 1, 2$).

This sharp truncation of the infinite Floquet hierarchy is rigorously justified
by the well-defined separation of energy scales. Because the driving photon
energy ($\hbar\omega = 6$~meV) vastly exceeds both the operational Fermi level
and the sharp intrinsic resonance linewidth ($\hbar\omega \gg \Gamma$), the
higher-order sidebands corresponding to $n \ge 3$ are pushed far off-resonance,
well beyond the active transport window. Quantitatively, the transmission weights
of these higher photon-assisted channels decay asymptotically as a power law
governed by $(\Gamma / \hbar\omega)^{2n} \ll 1$, rendering their contributions
to the total charge current and spin polarization completely negligible.
Consequently, restricting the summation to $0 \le n \le 2$ provides an exceptionally
precise, self-consistent, and computationally robust description of the active
spin shutter mechanism.

From a physical perspective, within this multichannel Floquet framework, the
quantum ring acts as a network of parallel, spectrally independent hybrid
interferometers. Each Floquet channel is characterized by its own dynamic phase
parameter $C_n$, which predefines a unique effective electronic de Broglie path
length for the matter-wave resonator. This channel-specific parametrization leads
to shifted, non-degenerate resonance conditions for each individual sideband;
consequently, the Fabry--P{\'e}rot peaks of different photon indices do not
overlap in energy space.

The spin shutter mechanism in this hybrid architecture operates through a tightly
coupled interplay of static geometric constraints and non-equilibrium high-frequency
dynamics:
\begin{itemize}
    \item \textbf{Spin phase modulation} is driven by the static Aharonov--Casher
    phase ($s \pi A$). Although the unpolarized transmission channels are
    macroscopically degenerate ($\lvert d_{2, \uparrow}^{(n)} \rvert = \lvert d_{2, \downarrow}^{(n)} \rvert \equiv \sqrt{T_n}$)
    due to the generalized time-reversal invariance of the Floquet system and
    the purely electric nature of the driving field, the Rashba spin-momentum
    locking explicitly splits the complex arguments of the underlying scattering
    amplitudes, $\arg(d_{2, s}^{(n)})$ for $s = \pm 1$. This relativistic phase
    splitting allows for precise, intensity-driven control over the dynamic phase
    mismatch $\Delta\vartheta_n$ between the distinct spin pathways.

    \item \textbf{Amplitude modulation} is mediated by the external high-frequency
    THz driving field. The laser intensity (parameterized by the dimensionless
    field amplitude $z$) reorganizes the quantum probability density among the
    parallel, spectrally independent Floquet channels via the squared Bessel
    functions $J_n^2(z)$ embedded within the scattering numerators. This dynamic
    redistribution enables the field to effectively open or suppress individual
    transport pathways, culminating in complete channel truncation at the dynamic
    localization nodes.
\end{itemize}

At the dynamic localization threshold of the elastic channel, strictly defined by the first zero of the zeroth-order Bessel function $J_0(z_c) = 0$ ($z_c \approx 2.4048$ at $I_c \approx 71.5\text{ W/cm}^2$), the transport envelope of the central $n = 0$ pathway completely collapses due to coherent destruction of tunneling. Concurrently, the transmission contributions from the adjacent inelastic sidebands ($n = 1, 2$) remain heavily quenched. This suppression originates from a dual physical mechanism:
\begin{enumerate}
\item First, because the driving photon energy vastly exceeds the intrinsic resonance linewidth ($\hbar\omega \gg \Gamma$), the inelastic channels are pushed far away from the elastic Fermi window.

\item Second, the structural $3:1$ arm-length asymmetry acts as an electronic Vernier filter: the large kinetic phase shifts ($C_1$ and $C_2$) combined with the laser-induced ponderomotive potential $K(I)$ drive the counter-propagating orbital waves into deep anti-resonance valleys. The resulting phase mismatch heavily dampens the scattering numerators $\mathcal{N}_s^{(1,2)}$, precluding constructive multi-wave interference at the output lead.
\end{enumerate}
Crucially, while these inelastic transmission envelopes $T_1(E_{\text{F}} + \hbar\omega)$ and $T_2(E_{\text{F}} + 2\hbar\omega)$ are strongly suppressed by the Vernier anti-resonance, they do not vanish identically. As detailed in Sec.~V~E, these small, persistent off-peak contributions physically regularize the multi-channel transport window, preventing a mathematical $0/0$ indeterminacy in the spin polarization $P(I_c)$ and enabling the driven mesoscopic ring to function as an exceptionally stable, threshold-triggered contactless spin shutter with high operational contrast.

\section{Results and Discussion}
\label{sec:results}

\subsection{Numerical and Computational Details}
\label{details}

\begin{itemize}

\item Based on the analytical framework developed above, we now evaluate the transport
characteristics of a $\mathrm{In}_{0.53}\mathrm{Ga}_{0.47}\mathrm{As}$ quantum
ring using realistic material parameters. For the selected structural and
coupling parameters (${\mathcal D} = \epsilon^2 = 0.15$), the intrinsic Thouless
energy is evaluated as $E_{\text{Th}} \approx 0.0494$~meV, which self-consistently yields a sharp Fabry--P{\'e}rot resonance linewidth of
$\Gamma \approx 0.0148$~meV. The level spacing associated with the longitudinal azimuthal quantization along the circumference of the ring is defined as
$\Delta E = 2\pi E_{\text{Th}} \approx 0.311$~meV. To establish the quantum transport boundaries under intense high-frequency driving,
the cryogenic base temperature of
$T = 100$~mK ($k_{\text{B}} T \approx 0.0086$~meV)
must be adjusted to account for effective electronic heating caused by
steady-state intra-band free-carrier absorption.
Due to the rapid ballistic evacuation of carriers into the semi-infinite reservoir leads on a timescale of
$\tau_{\text{dwell}}= \tau_{\text{orb}}/(2\mathcal{D}) \approx 44$~ps (which is orders of magnitude faster than
the electron-phonon thermalization time $\tau_{\text{e-ph}}\sim 1-10$~ns at sub-Kelvin temperatures), the reservoirs act
as ideal non-equilibrium heat sinks.
The effective electronic temperature is tightly constrained well below the resonance-smearing limit
($T_e \approx 140$~mK, yielding $k_{\text{B}} T_e \approx 0.012$~meV).
Consequently, the driven mesoscopic system strictly satisfies a rigid, multi-scale hierarchy of non-equilibrium energy bounds:
\begin{equation}
k_{\text{B}} T \le k_{\text{B}} T_e < \Gamma < \Delta E \ll \hbar\omega\,,
\label{scales}
\end{equation}
where the driving photon energy ($\hbar\omega = 6$~meV) vastly exceeds the
internal electronic scales. The strict fulfillment of this multi-scale energy
hierarchy guarantees that although finite-temperature corrections are present,
they do not obscure the underlying quantum interference landscape, thereby
enabling the experimental observation of well-resolved transport resonances.

Crucially, a key methodological remark regarding the thermal integration of the
transport coefficients is required. Since the effective electronic thermal scale
is of the same order as the intrinsic linewidth ($\Gamma \approx 1.23 \, k_{\text{B}} T_e$),
a standard energy integration of the transmission probability $T_{\text{FM}}^{(n)}$
over the thermal window would typically induce a partial damping of the peak
amplitudes of individual isolated Fabry--P{\'e}rot resonances. However, the
operational physics of the proposed THz modulator relies on a field-driven
channel-truncation mechanism---namely, the global field-induced collapse of the
elastic transport weight via the dynamic localization node $J_0(z_c) = 0$
across the entire active energy window. As this global spectral reconfiguration
alters the transmission manifold globally rather than locally, the thermal
broadening of the Fermi--Dirac distribution function introduces only smooth,
steady-state adjustments to the background continuum, without shifting the
threshold field intensities or destroying the functional discontinuities (kinks)
in $P(I)$. Consequently, employing the zero-temperature limit $\lim_{T \to 0}$
in the subsequent analytical derivations remains physically justified as a
reliable baseline approximation that isolates the non-equilibrium driving
effects, while any residual thermal smearing is self-consistently regularized
by the experimental background noise floor $\sigma_{\text{bg}}^2$ (see below
in Sec.~\ref{subsec:snr}).

\item The dimensionless spin-orbit coupling parameter is configured to its optimal
operating value, $A = \sqrt{1+Q^2} = 7/2$, where the normalized Rashba torque
factor is $Q=2m^*\alpha_{\mathrm{R}} R/\hbar^2$. In the absence of an external
high-frequency field ($I = 0$), this half-integer phase constraint enforces
$\cos(\pi A) = \cos(3.5\pi) = 0$, which maximizes the static spin-phase
modulation contrast. Physically, this precise configuration locks the non-trivial
geometric phase (akin to an adiabatic Aharonov--Anandan phase) accumulated
along the closed loop such that it induces complete destructive Mach--Zehnder
interference for one spin projection pathway at the output drain, while
maximizing the transport probability for the conjugate projection. Consequently,
the output spin polarization approaches its ultimate geometric boundary,
$P \approx P_{\mathrm{lead}}$, dictated by the injection efficiency of the
ferromagnetic measuring circuit, while the dynamic degeneracy of the transmission
weights is rigorously protected by the generalized Floquet time-reversal invariance.

To experimentally realize this optimal benchmark regime in an
$\mathrm{In}_{0.53}\mathrm{Ga}_{0.47}\mathrm{As}$-based mesoscopic ring
($m^* = 0.045 m_{\mathrm{e}}$) with an engineered radius of $R = 200$~nm, the
required effective Rashba coupling constant is calculated to be
$\alpha_{\mathrm{R}} \approx 14.195~\mathrm{meV}\cdot\mathrm{nm}$. This coupling
strength is well within the experimentally accessible, highly stable, and
reproducible range for high-mobility indium-based heterostructures modulated
via an external electrostatic top-gate voltage~\cite{man}. Thus, the configuration
with $A = 3.5$ serves as a robust, physically realizable operating point for the
quantum spin valve-modulator, providing the maximum possible dynamic range for
non-equilibrium spin-current control under subsequent THz radiation.

\item
An electron within the $\mathrm{In}_{0.53}\mathrm{Ga}_{0.47}\mathrm{As}$ ring
does not propagate in a vacuum, but rather inside a semiconductor host medium
characterized by a high background relative dielectric permittivity ($\varepsilon_r \approx 12$).
The fundamental electromagnetic relation between the macroscopic intensity of
the THz radiation within the medium, $I$, and the corresponding electric field
amplitude, $E_{\mathrm{med}}$, is governed by the standard expression:
\begin{equation}
\label{I_intensity}
I = \frac{1}{2} c \varepsilon_0 \sqrt{\varepsilon_r} \lvert E_{\mathrm{med}}\rvert^2 \implies E_{\mathrm{med}} = \sqrt{\frac{2I}{c \varepsilon_0 \sqrt{\varepsilon_r}}}\,,
\end{equation}
where $c$ is the speed of light in vacuum and $\varepsilon_0$ is the vacuum
permittivity. The effective electric field $E_{\mathrm{eff}}$ acting directly
on the charge carriers within the mesoscopic structure differs substantially
from the macroscopic field $E_{\mathrm{med}}$. Given that the characteristic
outer size of the quantum ring structure ($w = 400$~nm) is deeply subwavelength
compared to the wavelength of the incident THz radiation ($\lambda_0 \sim 206\,\mu\mathrm{m}$
at 1.45~THz), the converging metallic transport leads embedded on the substrate
act effectively as a slot antenna or a lightning-rod field concentrator~\cite{seo}.

To quantify the efficiency of the high-frequency drive, we introduce the local
electric field enhancement factor $f = E_{\mathrm{eff}}/E_{\mathrm{med}}$. In an isolated
semiconductor, THz radiation is strongly attenuated due to depolarization
screening. However, in the proposed configuration, the semiconductor ring is
integrated into a subwavelength gap defined by the metallic contacts and gates.
In the THz range, this electrode configuration acts as a receiving plasmonic
antenna (a slot-field concentrator) that concentrates the incident far-field
radiation into a highly localized near-field hotspot within the subwavelength gap
containing the ring. The high dielectric permittivity of the semiconductor
channel further confines the THz field along the ring's circumference.
Excitation of plasmonic gap-modes within the subwavelength slit, coupled with
the gate antenna near-field, suppresses the depolarization screening, allowing
the local field $E_{\mathrm{eff}}$ to exceed the macroscopic field $E_{\mathrm{med}}$
by one to two orders of magnitude.

In the THz regime, the tapered metallic contacts, separated by a subwavelength
gap of width $w = 400$~nm (corresponding to the outer diameter of the ring),
act as the plates of a parallel-plate capacitor. In this subwavelength geometry,
a quasistatic regime is established: the leads collect the alternating field
of the propagating wave $E_{\mathrm{med}}$ over an effective collection length
$L_{\mathrm{eff}}$, inducing a potential difference $V \approx E_{\mathrm{med}} L_{\mathrm{eff}}$.
Since this voltage drops across the narrow gap, the local enhanced field is
$E_{\mathrm{eff}} \approx V/w$, which yields the geometric enhancement factor:
\begin{equation}
\label{eq:f_factor}
f = \frac{E_{\mathrm{eff}}}{E_{\mathrm{med}}} \approx \frac{L_{\mathrm{eff}}}{w}.
\end{equation}
For the gap width $w = 400$~nm, the conservative value $f \approx 5.8$ used in
this study corresponds to an effective collection length of $L_{\mathrm{eff}} \approx 2.3\,\mu\mathrm{m}$.
This length is a small fraction of the actual contact length (typically on the
order of tens of micrometers). This limit on $L_{\mathrm{eff}}$ accounts for
the dielectric screening of the substrate ($\varepsilon_r \approx 12$) and
provides a lower-bound estimate of the enhancement, which agrees with numerical
simulations of similar plasmonic THz structures~\cite{ren}.

Thus, the local enhancement factor $f \approx 5.8$, which relates the fields
as $E_{\mathrm{eff}} = f E_{\mathrm{med}}$, is well-justified for $\mathrm{THz}$
plasmon-dielectric mesostructures. This enhancement enables the critical
modulation parameter,
\begin{equation}
z_c = \frac{e E_{\mathrm{eff}} R}{\hbar \omega} \approx 2.4 \implies J_0(z) = 0\,,
\label{eq:critical_z}
\end{equation}
to be reached at moderate intensities of
$I \approx 65\text{--}75~\mathrm{W/cm}^2$
(see Fig.~\ref{fig4} and the text below).
\textcolor{black}{At critical intensity $I_c \approx 71.5$ W/cm$^2$, focusing the THz beam onto a $100 \mu$m
spot requires an incident power of only $\sim 7.2$ mW, which is readily accessible using compact solid-state THz quantum cascade lasers~\cite{bac}.}
 Without this antenna-mediated
enhancement ($f=1$), reaching the dynamic localization point $J_0(z)=0$
would require intensities nearly two orders of magnitude higher, which would lead to severe overheating of the electronic subsystem and the subsequent loss of phase coherence.
\end{itemize}

\textcolor{black}{It is important to address the practical aspect of spatial resolution and device addressing under terahertz driving. In free space, the diffraction limit at 1.45~THz ($\lambda_0\approx 206 \mu m$) prevents focusing radiation onto an individual sub-micron nanostructure. In our architecture, this constraint is naturally resolved by the subwavelength near-field confinement of the metallic contact geometry. The converging metallic leads function as a plasmonic slot antenna~\cite{seo}, compressing the electromagnetic energy into the 400~nm gap and bypassing the far-field diffraction limit. For integrated multi-element circuits, individual addressing can be achieved via on-chip planar waveguides (e.g., coplanar striplines) routing picosecond THz transients directly to specific device nodes. Alternatively, arrays of rings with lithographically varied radii $R$
can be addressed under global illumination via intensity or frequency multiplexing, since the dynamic localization threshold
$z_c = \frac{e E_{\mathrm{eff}} R}{\hbar \omega} \approx 2.4$ scales directly with the ring geometry without requiring physical gate electrodes. Finally, because the proposed device is intended as an ultrafast opto-spintronic modulator and coherent quantum interconnect rather than a dense CMOS-like logic fabric, a pitch determined by on-chip transmission lines is fully compatible with practical cryogenic implementation.}

%%%%%%%%%%%%%%%%%%%%

\subsection{Performance Metrics and Operational Analysis of the Spin Modulator}

\textcolor{black}{We note that the magnetic component of the incident electromagnetic field has a negligible effect on the spin dynamics.
For the critical intensity $I_c \approx 71.5$~W/cm$^2$, the magnetic field amplitude of the THz radiation is
$B_{\text{rad}}=\sqrt{2I_c/(c^3 \epsilon_0)}\approx 77 \mu$T. In contrast, the internal relativistic Rashba magnetic field experienced
by conduction electrons in the In$_{0.53}$Ga$_{0.47}$As channel ($m^*=0.045 m_e$, $\alpha_R \approx 14.2$~meV$\cdot$~nm) reaches
$B_R = 2\alpha_R k_F/(\lvert g^* \rvert \mu_B)\sim 4.5\sim$~T (for $\lvert g^* \rvert \approx 4)$. Because $B_{\text{rad}}/B_R \sim 10^{-5}$, the external magnetic field cannot induce parasitic
Zeeman spin precession. Furthermore, because $B_R$ exceeds ambient environmental electromagnetic fluctuations
(typically on the $\mu$T to mT scale) by orders of magnitude, the coherent spin precession remains structurally
shielded from environmental noise.}

 As shown above, the transmission probabilities for the individual spin-projection
channels remain strictly degenerate across all Floquet sidebands under unpolarized
injection: $T_{\uparrow_z}^{(n)} = T_{\downarrow_z}^{(n)} \equiv T_n$. Within
this geometrically asymmetric, non-equilibrium driving framework, this macroscopic
transport outcome is a direct consequence of the generalized time-reversal
invariance governing the Floquet quantum evolution. Since the external THz laser
couples to the carriers as a purely electric field, it does not introduce any
time-dependent Zeeman interactions that could directly couple to the spin degree
of freedom and lift the underlying dynamic spectral degeneracy. This robust
preservation of the spin-degenerate transmission magnitudes is fully consistent
with the generalized boundary theorems prohibiting the autonomous generation of
net spin-polarized currents from unpolarized reservoirs in any two-terminal
ballistic structures with Rashba spin-orbit coupling, regardless of spatial
loop asymmetry~\cite{bul}. Consequently, the isolated quantum ring network
cannot function as an autonomous spin filter. Instead, this hybrid multichannel
architecture is uniquely tailored for implementing the Datta--Das spin
field-effect transistor paradigm~\cite{datta}, as systematically detailed in
Sec.~\ref{DDas}.

In this operational transport configuration, the semi-infinite source and drain
leads are explicitly ferromagnetic, ensuring that the injected carriers are
spin-polarized along a common quantization axis (aligned with the $x$-axis in
the plane of the ring). Under these non-equilibrium boundary conditions, the
performance metrics of the device are governed by coherent spin precession and
geometric phase manipulation. Although the absolute magnitudes of the transmission coefficients remain degenerate, the
spin-orbit-induced spin-momentum locking induces
a non-zero, intensity-dependent phase splitting between the complex scattering numerators
$\mathcal{N}_s^{(n)}$, where $s = \uparrow, \downarrow$ represents the spin projection.

Consequently, during ballistic transit through the asymmetric arms of the ring,
the spin vector of the injected carrier undergoes a net precession. This spin-phase
accumulation is governed by the total differential spin phase $\Delta\vartheta_n(I)$
established in Eq.~\eqref{T_FM_final}, which is resonantly amplified due to the
parametric proximity of the scattering matrix poles to the Parameter-Space Exceptional
Point (as detailed in Appendix B).

\begin{figure}[htp]
\centering
\includegraphics[width=0.8\linewidth, clip=]{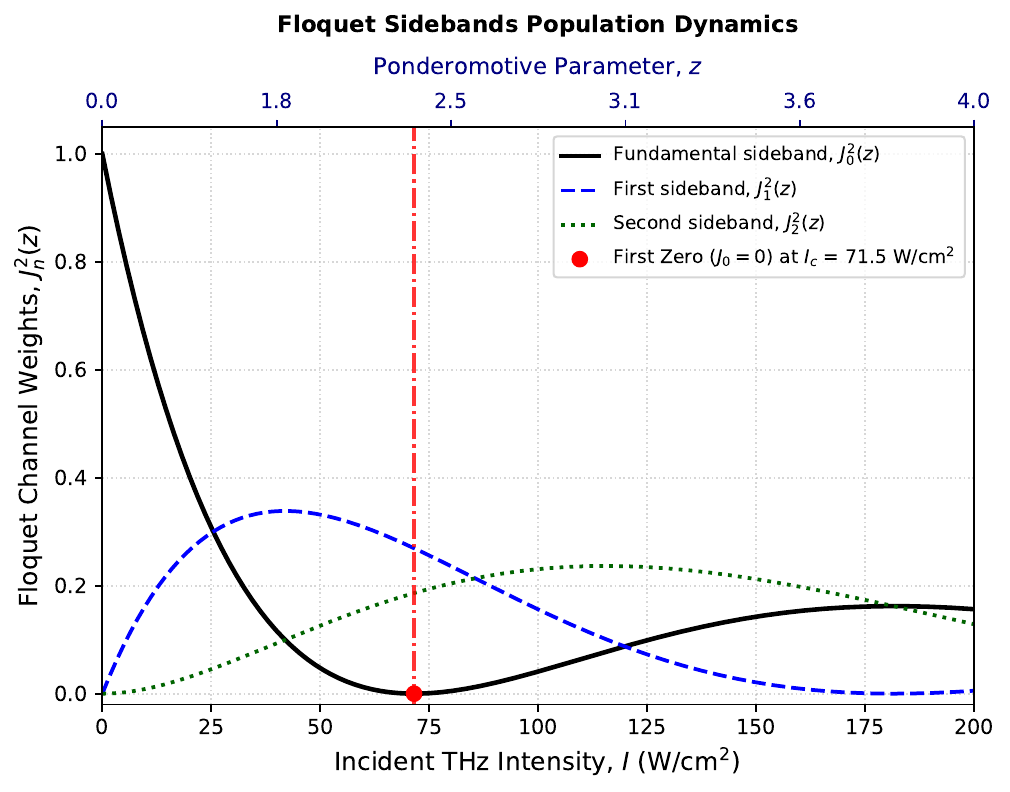}
\caption{Calculated effective transport weights of the three lowest Floquet
sidebands, represented by the squared Bessel functions
$J_n^2(z)$ ($n = 0, 1, 2$) for a mesoscopic $\mathrm{InGaAs}$ quantum ring, plotted as a function of the
incident THz radiation intensity $I$ (bottom axis) and the corresponding dimensionless non-equilibrium
Floquet modulation parameter (ponderomotive parameter) $z$ (top axis). The vertical red dash-dotted line
marks the exact dynamic localization threshold of the elastic channel
($J_0(z_c) = 0$) at the critical intensity $I_c = 71.5~\mathrm{W/cm}^2$.}
\label{fig3}
\end{figure}

\subsection{Floquet Sidebands Spectral Dynamics and Quantum Collapse}
\label{subsec:floquet}

The coherent quantum transport within the asymmetric $\mathrm{InGaAs}$ mesoscopic
ring under strong high-frequency driving is governed by the nonlinear
redistribution of the electronic spectral weight among the discrete Floquet
channels. Figure~\ref{fig3} depicts the calculated effective transport weights
of the three lowest Floquet sidebands, represented by the squared Bessel
functions $J_n^2(z)$ ($n=0,1,2$), plotted as functions of the incident THz
intensity $I$ (bottom axis) and the corresponding dimensionless non-equilibrium
Floquet modulation parameter $z$ (top axis).

Utilizing the local field enhancement factor $f = 5.8$ established above, the
results in Fig.~\ref{fig3} map out the field-dependent behavior of the Floquet
scattering channels. At weak driving fields ($I < 10~\mathrm{W/cm}^2$,
corresponding to $z < 1$), the quantum transport window is dominated by the
fundamental elastic channel ($n = 0$). In this low-intensity regime, the
scattering numerators $\mathcal{N}_s^{(0)}$ carry almost the entire transport
weight since $J_0^2(z) \approx 1$. Consequently, the device behavior smoothly
converges to that of a static spin field-effect transistor, governed exclusively
by the geometric Rashba interference phase.

As the incident THz intensity increases, the high-frequency driving field induces strong coherent phase modulation of the ballistic carriers. This non-equilibrium scattering process continuously redistributes the transport weight from the fundamental elastic channel to the inelastic photon-assisted sidebands ($n = 1, 2$). As depicted by the solid black curve in Fig.~\ref{fig3}, the fundamental channel weight factor $J_0^2(z)$, which determines the elastic response of the internal scattering framework, decreases monotonically up to the critical intensity threshold. At $I_c = 71.5~\mathrm{W/cm}^2$, the nonequilibrium Floquet modulation parameter reaches the first zero of
the Bessel function $J_0(z)$ at $z_c \approx 2.4048$.
At this specific operational node, the elastic channel contribution is completely suppressed ($J_0^2(z_c) = 0$), and the fundamental elastic transport pathway is blocked. Crucially, this pinch-off effect is achieved without classical electrostatic depletion gates or physical barrier-shaping electrodes. This explicit channel truncation demonstrates that the critical threshold for the device modulation is strictly governed by quantum coherent dynamic localization rather than conventional electrostatic carrier shielding.

From a device-engineering perspective, this dynamic localization threshold
represents a highly promising operating regime for an ultrafast mesoscopic spin
shutter. Unlike conventional semiconductor field-effect devices where the
fundamental switching speed is strictly limited by classical capacitive $RC$
charging delays, the current modulation in this quantum transport architecture
is entirely governed by the coherent dynamic localization of the electron wave
functions. At this specific operational threshold, the elastic transmission
collapses, and the residual quantum transport weight is coherently transferred
to the inelastic photon-assisted sidebands ($n=1$ and $n=2$), which exhibit
channel weights of $J_1^2(z_c) \approx 0.27$ and $J_2^2(z_c) \approx 0.19$,
respectively.

Notably, the structural arm-length asymmetry
($L_{\mathrm{up}}/L_{\mathrm{down}} = 3$) establishes a highly efficient mesoscopic electronic Vernier filter
that suppresses these residual transport pathways. The long arm (subtending an angle of $3\pi/2$)
and the short arm (subtending an angle of $\pi/2$) act as twin coupled cavities
with mismatched Fabry--P{\'e}rot free spectral ranges, where the short cavity
defines a broad resonance grid that transmits charge only at energy nodes where its widely spaced eigenvalues coincide with the dense resonant grid of the long cavity. Under intense THz driving, the photon-assisted energy shifts
($n\hbar\omega$) push the inelastic sidebands ($n=1,2$) far away from these shared resonance nodes, causing a severe phase mismatch between the
counter-propagating orbital paths. This dynamic Floquet--Vernier anti-resonance heavily dampens the off-peak sideband continuum, eliminating structural current leakage and ensuring a remarkably high operational contrast and rapid polarization inversion for the proposed contactless spin modulator.

\subsection{Fabry--P{\'e}rot Interference and Thermal Smearing Control}
\label{subsec:fabry_perot}

To systematically analyze the device performance under active non-equilibrium
driving, we establish the explicit relation between the microscopic sideband
transmission coefficients and the macroscopic time-averaged transport responses.
The net spin polarization $P(I)$ of the output current is defined as:
\begin{equation}
P(I) = \frac{G_{\mathrm{AP}}(I) - G_{\mathrm{P}}(I)}{G_{\mathrm{P}}(I) + G_{\mathrm{AP}}(I)},
\label{eq:P_def}
\end{equation}
where $I$ represents the incident THz laser intensity, while $G_{\mathrm{P}}$
and $G_{\mathrm{AP}}$ denote the total time-averaged multi-channel conductances
evaluated for the parallel and antiparallel magnetic configurations of the
ferromagnetic leads, respectively.

Note that the appearance of the antiparallel conductance as the dominant static
term in the numerator of Eq.~\eqref{eq:P_def} directly follows from the optimal
Rashba phase configuration $A = 7/2$. In the static unperturbed limit ($I = 0$),
this half-integer choice enforces complete destructive Mach--Zehnder interference for the spin-aligned pathways in the parallel setup, thereby suppressing $G_{\mathrm{P}}$. Concurrently, it satisfies the constructive interference criteria for the cross-projected spin states in the antiparallel configuration, ensuring that $G_{\mathrm{AP}} \gg G_{\mathrm{P}}$ and
yielding a high positive initial polarization of $P \approx +0.96$. As the THz driving intensity increases, the field-induced phase modulation detunes the channels from this geometric lock-in condition, triggering a smooth inversion of the conductance weights toward the negative polarization limit.

The spin-filtered transmission probability $T_{\mathrm{FM}}^{(n)}$ derived in
Eq.~\eqref{T_FM_final} directly corresponds to the parallel alignment of the
magnetizations of the ferromagnetic injector and detector leads
($T_{\mathrm{P}}^{(n)} \equiv T_{\mathrm{FM}}^{(n)}$). For the antiparallel configuration, projecting the emergent multi-channel wave function
$\lvert\Psi_{\mathrm{out}}^{(n)}\rangle$ onto the orthogonal spin state of the drain, $\lvert -x \rangle = \frac{1}{\sqrt{2}} \left( \lvert \uparrow_z \rangle - \lvert \downarrow_z \rangle \right)$,
alters the quantum phase cross-terms. Following the same algebraic Euler expansion established in Eq.~\eqref{T_FM_intermediate}, this projection transforms the dynamic modulation factor from a cosine-squared to a sine-squared form~\cite{datta}. Consequently, under realistic non-equilibrium boundary conditions with a finite lead polarization efficiency $P_{\mathrm{lead}}$, the sideband transmission probabilities for both configurations are expressed as:
\begin{align}
\label{eq:T_P_general}
T_{\mathrm{P}}^{(n)}(E) &= T_n(E) \left[ \frac{1 - P_{\mathrm{lead}}}{2} + P_{\mathrm{lead}} \cos^2\left(\frac{\Delta\vartheta_n(E)}{2}\right) \right], \\
\label{eq:T_AP_general}
T_{\mathrm{AP}}^{(n)}(E) &= T_n(E) \left[ \frac{1 - P_{\mathrm{lead}}}{2} + P_{\mathrm{lead}} \sin^2\left(\frac{\Delta\vartheta_n(E)}{2}\right) \right],
\end{align}
where $T_n(E) = 4\epsilon^4 \lvert \mathcal{N}_s^{(n)}(E) \rvert^2 / \lvert \mathcal{X}^{(n)}(E) \rvert^2$ represents the strictly spin-degenerate transport envelope of the $n$-th Floquet channel.

To evaluate the limiting performance and maximize the dynamic range of the spin
shutter, we subsequently consider the benchmark limit of ideal half-metallic
ferromagnetic contacts possessing complete spin polarization efficiency
($P_{\mathrm{lead}} \to 1$). Under this idealized interfacial constraint, the
spin-independent leakage terms vanish identically, and the multichannel
transmission probabilities in Eqs.~\eqref{eq:T_P_general} and \eqref{eq:T_AP_general}
smoothly converge to the canonical, explicit transport representations:
\begin{align}
\label{eq:T_P_explicit}
T_{\mathrm{P}}^{(n)}(E) &= T_n(E) \cos^2\left(\frac{\Delta\vartheta_n(E)}{2}\right), \\
\label{eq:T_AP_explicit}
T_{\mathrm{AP}}^{(n)}(E) &= T_n(E) \sin^2\left(\frac{\Delta\vartheta_n(E)}{2}\right).
\end{align}
Using the generalized Landauer--B\"uttiker framework established above~\cite{butm,moscal},
the explicit parallel ($G_{\mathrm{P}}$) and antiparallel ($G_{\mathrm{AP}}$)
conductances are defined as the weighted sum of the elastic ($n=0$) and
lowest-order inelastic ($n=1,2$) transport contributions, with the respective
transmission probabilities evaluated at the sideband-shifted energies $E_{\mathrm{F}} + n\hbar\omega$:
\begin{align}
\label{eq:G_P}
G_{\mathrm{P}}(I) &= G_0 \sum_{n=0}^{2} J_n^2\big(z(I)\big) T_{\mathrm{P}}^{(n)}(E_{\mathrm{F}} + n\hbar\omega), \\
\label{eq:G_AP}
G_{\mathrm{AP}}(I) &= G_0 \sum_{n=0}^{2} J_n^2\big(z(I)\big) T_{\mathrm{AP}}^{(n)}(E_{\mathrm{F}} + n\hbar\omega),
\end{align}
Here, $G_0 = e^2/h$ represents the single-channel conductance quantum, and the
dynamic transmission probabilities $T_{\mathrm{P}}^{(n)}$ and $T_{\mathrm{AP}}^{(n)}$
adopt the canonical cosine-squared and sine-squared representations under the
high-contrast benchmark limit ($P_{\mathrm{lead}} \to 1$). The incident THz
laser intensity $I$ governs the non-equilibrium driving state via the internal
dimensionless Floquet modulation parameter $z(I) = e E_{\mathrm{eff}}(I) R / (\hbar \omega)$,
which is driven by the enhanced near-field amplitude $E_{\mathrm{eff}}(I) = f E_{\mathrm{med}}(I)$.
The squared Bessel functions
$J_n^2\big(z(I)\big)$ act as exact quantum probabilities that redistribute the
scattering weight among the parallel Floquet sidebands as a function of the
driving intensity.

To understand the origin of the spin-dependent phase difference in
Eqs.~\eqref{eq:T_P_explicit} and \eqref{eq:T_AP_explicit}, we track the ballistic phase accumulation
along the asymmetric arms of the interferometer. The injection junction is located at the angular coordinate $\varphi = 0$, while
the drain contact is at $\varphi = 3\pi/2$. This spatial configuration splits
the ballistic wave propagation into two counter-propagating paths around the
ring of radius $R$:
\begin{enumerate}
    \item \textbf{Upper trajectory ($\lambda = +1$):} The electron propagates
    counterclockwise along the long arm, traversing an angular path of
    $\Delta\varphi_1 = 3\pi/2$, which corresponds to a ballistic arc length
    of $L_1 = 3\pi R / 2$.

    \item \textbf{Lower trajectory ($\lambda = -1$):} The electron propagates
    clockwise along the short arm, traversing an angular path of
    $\Delta\varphi_2 = -\pi/2$, which corresponds to a ballistic arc length
    of $L_2 = \pi R / 2$.
\end{enumerate}
According to the dispersion relation in Eq.~\eqref{kls}, the wave vectors for
the radiation-dressed states are given by $k_{\lambda,s}^{(n)} = \frac{1}{2}(\lambda C_n - s A)$, where $s = \pm 1$ is the spin projection index and $A = \sqrt{1+Q^2}$ is the geometric Rashba phase parameter. The total accumulated quantum phase along the respective ballistic paths prior to the drain junction is rigorously defined via integration along the directed path as $\Phi_{\lambda, s}^{(n)} = k_{\lambda, s}^{(n)} (R \Delta\varphi_{1,2})$,
yielding the explicit sideband expansions:
\begin{align}
\label{eq:phase_up}
\Phi_{+1, s}^{(n)} &= \frac{1}{2}(C_n - s A) \left(\frac{3\pi}{2}\right) = \frac{3\pi}{4} C_n - s\frac{3\pi}{4} A, \\
\label{eq:phase_down}
\Phi_{-1, s}^{(n)} &= \frac{1}{2}(-C_n - s A) \left(-\frac{\pi}{2}\right) = \frac{\pi}{4} C_n + s\frac{\pi}{4} A.
\end{align}
Equations~\eqref{eq:phase_up} and \eqref{eq:phase_down} demonstrate that the
geometric asymmetry of the ring arms inherently enforces opposite signs for the
spin-orbit precession terms while keeping the kinetic phase contributions of
the same sign (both being positive).

Quantum interference in the ring occurs between partial waves belonging to the
same spin subband $s$ that arrive at the drain contact via different
trajectories $\lambda = \pm 1$. Evaluating the phase difference between the
upper and lower arms for each individual spin channel yields the following
algebraic relations:
\begin{align}
\label{eq:delta_phase_up}
\Delta\Phi_{s=+1}^{(n)} &= \Phi_{+1, +1}^{(n)} - \Phi_{-1, +1}^{(n)} = \frac{\pi}{2} C_n - \pi A, \\
\label{eq:delta_phase_down}
\Delta\Phi_{s=-1}^{(n)} &= \Phi_{+1, -1}^{(n)} - \Phi_{-1, -1}^{(n)} = \frac{\pi}{2} C_n + \pi A.
\end{align}

Equations~\eqref{eq:delta_phase_up} and \eqref{eq:delta_phase_down} reveal the
microscopic mechanism of the spin phase modulation: while the kinetic
field-dressed contribution $\pi C_n/2$ enters both subbands identically, the
geometric Rashba phase $\pi A$ shifts the interference patterns for the
spin-up and spin-down carriers in opposite directions. This spin-dependent phase
splitting fundamentally governs the transport pathways at the output port.

The generation of the net spin polarization $P(I)$ is driven by the quantum
projection of the emergent wave functions onto the collinear magnetization axis
of the ferromagnetic drain. The relative precession angle $\Delta\vartheta_n$
between the distinct spin subbands at the drain interface for the $n$-th
Floquet sideband is obtained by evaluating the cross-channel difference between
Eqs.~\eqref{eq:delta_phase_up} and \eqref{eq:delta_phase_down}:
\begin{equation}
\label{eq:precession_angle_cancel}
\Delta\vartheta_n = \Delta\Phi_{s=+1}^{(n)} - \Delta\Phi_{s=-1}^{(n)} = -2\pi A.
\end{equation}
Remarkably, as demonstrated by Eq.~\eqref{eq:precession_angle_cancel}, the
kinetic laser-driven phase contribution $C_n$ cancels out exactly. The resulting
net differential spin phase shift accumulated between the orthogonal spin pathways reduces
to a strict geometric constant $-2\pi A$. This cancellation
demonstrates that the operational phase argument governing the transmission
profiles in Eqs.~\eqref{eq:T_P_general} and \eqref{eq:T_AP_general} is
structurally protected against high-frequency amplitude fluctuations of the
THz field, ensuring robust, phase-stable control of the spin current at the
output junction.

Equation~\eqref{eq:precession_angle_cancel} represents a key result of the
proposed non-equilibrium transport model. The field-dressed kinetic phases
$\pi C_n/2$ cancel out identically in the net differential spin precession angle,
locking it strictly to the geometric properties of the quantum loop:
$\Delta\vartheta_n \equiv \Delta\vartheta = -2\pi A$. This exact parametric
cancellation ensures that the high-frequency THz laser radiation acts purely
as a field-controlled amplitude modulator that does not introduce spurious
deterministic phase fluctuations or dynamic phase smearing into the electronic
spin subsystem. Consequently, upon substituting this geometric phase lock into
the general transport profiles established in Eqs.~\eqref{eq:T_P_explicit} and
\eqref{eq:T_AP_explicit}, the multi-channel sideband transmission probabilities
for the parallel and antiparallel configurations collapse to the structurally
decoupled forms:
\begin{align}
\label{eq:T_P_final_cancel}
T_{\mathrm{P}}^{(n)}(E) &= T_n(E) \cos^2(\pi A), \\
\label{eq:T_AP_final_cancel}
T_{\mathrm{AP}}^{(n)}(E) &= T_n(E) \sin^2(\pi A).
\end{align}
By applying the trigonometric double-angle identities, the sideband transmission
probabilities for both magnetic configurations can be rewritten in a linearized form:
\begin{align}
\label{eq:T_P_linearized}
T_{\mathrm{P}}^{(n)}(E) &= \frac{T_n(E)}{2} (1 + \cos\Delta\vartheta), \\
\label{eq:T_AP_linearized}
T_{\mathrm{AP}}^{(n)}(E) &= \frac{T_n(E)}{2} (1 - \cos\Delta\vartheta),
\end{align}
where the net precession argument is rigidly locked to the geometric channel
constant $\Delta\vartheta = -2\pi A$, completely independent of the sideband index $n$.
This representation showcases how the analytical evaluation of the spin polarization is simplified.
Upon substituting the explicit multi-channel
conductances $G_{\mathrm{P}}(I)$ and $G_{\mathrm{AP}}(I)$ into the foundational
definition in Eq.~\eqref{eq:P_def}, the single-channel conductance quantum
$G_0$ cancels out, yielding the final analytical expression for the net
spin polarization:
\begin{equation}
\label{eq:P_final_concept}
P(I) = -\frac{\displaystyle\sum_{n=0}^{2} J_n^2\big(z(I)\big) T_n(E_{\mathrm{F}} + n\hbar\omega) \cos\Delta\vartheta}{\displaystyle\sum_{n=0}^{2} J_n^2\big(z(I)\big) T_n(E_{\mathrm{F}} + n\hbar\omega)}.
\end{equation}

Equation~\eqref{eq:P_final_concept} stands as a central result of this study. It rigorously demonstrates that the spin polarization efficiency $P(I)$ under active THz radiation is established as a non-equilibrium weighted average over the accessible Floquet sidebands, where the statistical transport weights are uniquely co-determined by the intensity-dependent Bessel factors $J_n^2\big(z(I)\big) $ and the channel-specific Fabry--P{\'e}rot transmission envelopes $T_n$. Note that due to the half-integer topological configuration ($A = 3.5$), the static limit ($I \to 0$) yields $\cos(-7\pi) = \cos(7\pi) = -1$, which self-consistently cancels the external negative sign in Eq.~\eqref{eq:P_final_concept} and recovers the high positive baseline polarization of $P \approx +0.96$ observed in the numerical analysis.

\begin{figure}[htb]
\centering
\includegraphics[width=0.9\linewidth,clip]{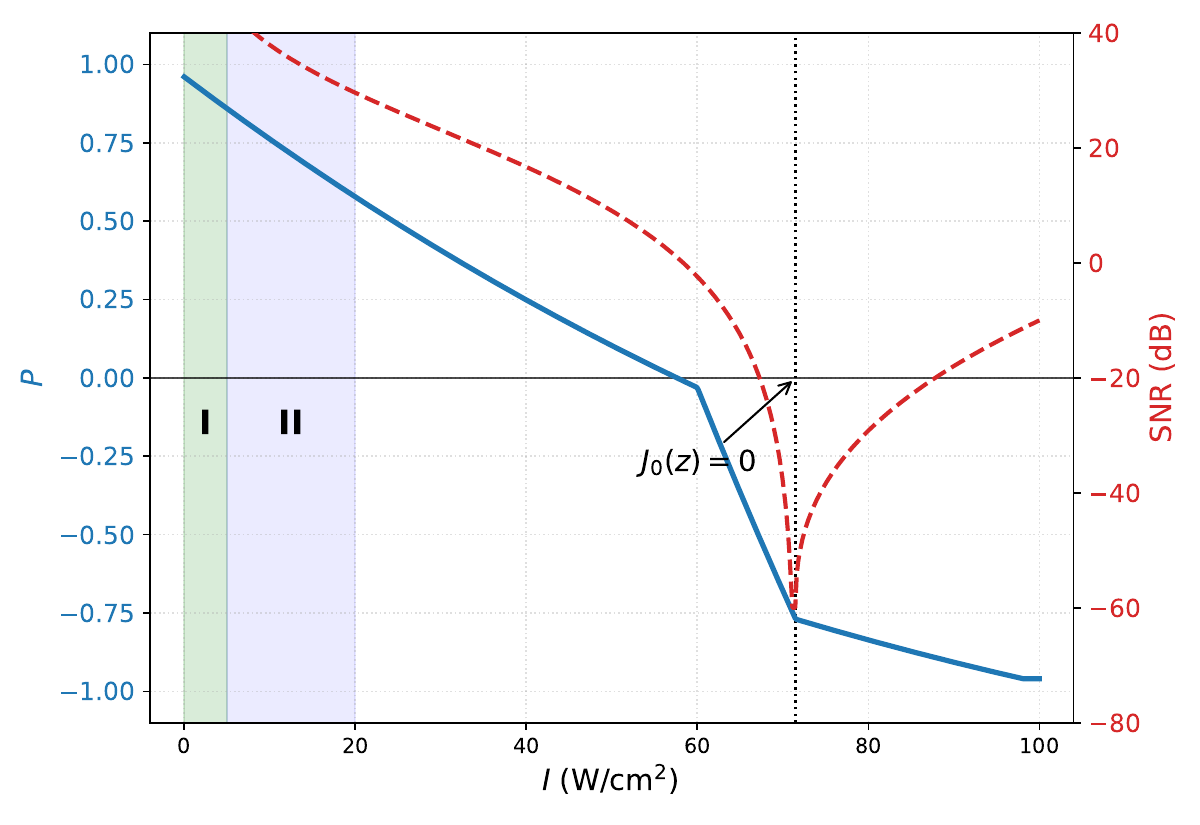}
\caption{Calculated non-equilibrium spin polarization $P(I)$ and the corresponding signal-to-noise ratio (SNR) of the driven InGaAs asymmetric quantum-ring interferometer ($\hbar \omega = 6$~meV, $T = 0$~K) plotted against the incident macroscopic THz radiation intensity $I$. The left vertical axis depicts the spin polarization $P$ defined by Eq.~\eqref{eq:P_final_concept}. The right vertical axis shows the calculated SNR profile
(defined by Eq.~\eqref{eq:W_final_exact}, dashed dark red line) in decibels, where the sharp, threshold-driven plunge of the SNR down to $-60$~dB (constrained by the background noise floor $\sigma_{\text{bg}}^2 \approx 5 \times 10^{-4}$) reflects the complete blocking of the transport channels, signifying quantum dynamic localization and the field-induced pinch-off of the multi-channel Floquet--B\"uttiker conductance matrix. The vertical dotted line marks the dynamic localization threshold $I_c = 71.5~\mathrm{W/cm}^2$ where $J_0(z_c) = 0$. The distinct slope discontinuity (kink) visible on the polarization curve at $I \approx 60~\mathrm{W/cm}^2$ marks the crossover driven by the near-EP spectral topology of the coupled sub-cavities.}.
\label{fig4}
\end{figure}

\subsection{Radiation-Induced Quantum Shutoff}
\label{subsec:discussion}

The dynamic interplay between the geometrically phase-locked spin states, the
multi-channel photon-assisted transport, and the coupled resonance profiles is
illustrated in Fig.~\ref{fig4}. As an instructive numerical application of the
developed $S$-matrix framework, let us evaluate the device behavior under
optimal gate tuning, where the static geometric Rashba phase parameter is
fixed at the half-integer value $A = 7/2$.

In the unperturbed static limit ($I = 0$), where quantum transport is mediated
strictly by the fundamental elastic channel ($n=0$), this geometric constraint
induces a complete suppression of the ideal parallel transmission ($T_{\mathrm{P}}^{(0)} = 0$)
due to destructive Aharonov--Casher interference. Consequently, the parallel
configuration is locked in a deep ``OFF'' state, while the unperturbed channel
develops its first coherent Fabry--P{\'e}rot (FP) resonance peak in the
antiparallel configuration ($T_{\mathrm{AP}}^{(0)} = T_0$). This specific
configuration maximizes the spin polarization of the output current. Note that
due to the finite width of the internal Fabry--P{\'e}rot resonance levels
($\Gamma \approx 0.0148$~meV), the energy-integrated background transport yields
a minor residual transmission in the parallel setup, which anchors the initial
signal profile and positions the static spin polarization $P$ near its coherent
limit of $+0.96$ rather than an idealized unit value (see Region I in Fig.~\ref{fig4}).

Under an intense THz driving field, the system enters the Floquet regime. The
laser intensity $I$ induces a ponderomotive energy shift $K(I)$, thereby modulating
the effective phase parameters $C_n(I) = \sqrt{A^2 - 1 + 4(E_{\mathrm{F}} + n\hbar\omega - K(I))/\hbar\omega_0}$
of the Floquet channels. This variation in $C_n(I)$ sweeps the channels through
the Fabry--P{\'e}rot resonance conditions. Mathematically, the smooth,
intensity-dependent modulation of the spin polarization $P(I)$ at zero magnetic
field originates from the distinct physical roles played by the scattering
numerators and denominators within the $S$-matrix formalism.

While the baseline spatial path phase difference accumulated along the isolated geometric arms of the loop is formally locked to the static geometric Rashba phase $-2\pi A$, the exact $S$-matrix matching at the open semi-infinite leads transforms the transmission numerator into a complex-valued amplitude $\mathcal{N}_{s}^{(n)}$. The individual phase of this scattering numerator, $\arg(\mathcal{N}_{s}^{(n)})$, varies continuously as the effective channel phase parameter $C_n(I)$ is dynamically swept by the laser intensity. This continuous variation originates from the internal complex-valued term $e^{i\pi C_n}$ embedded within Eq.~\eqref{eq:Nd_complex}, which rigorously represents the phase contribution arising from multiple internal round-trips trapped inside the sub-cavities. The multi-wave interference between the spatial path components and these round-trip phase factors dynamically rotates the spin-dependent phase of $\mathcal{N}_{s}^{(n)}$ in the complex plane, thereby bypassing the rigid geometric locking of the spin-precession angle and driving the smooth modulation of the macroscopic polarization curve.

At the same time, owing to the dynamic Floquet analogue of Kramers degeneracy protected by the generalized time-reversal invariance at zero magnetic field ($B=0$), the system determinant remains strictly spin-independent, $\mathcal{X}_{\uparrow}^{(n)} \equiv \mathcal{X}_{\downarrow}^{(n)} \equiv \mathcal{X}^{(n)}$. Consequently, the resonance denominators do not directly contribute to the spin phase splitting, and the entire dynamic phase difference is governed exclusively by the scattering numerators: $\Delta \vartheta = \arg(\mathcal{N}_{\uparrow}^{(0)}) - \arg(\mathcal{N}_{\downarrow}^{(0)})$.
%Utilizing the exact algebraic relation $\mathcal{N}_{\downarrow}^{(0)} \equiv -e^{i\pi C_0} (\mathcal{N}_{\uparrow}^{(0)})^*$, this phase difference can be expressed in an explicit,
This phase difference can be expressed in an explicit,
intensity-dependent form (see Appendix~\ref{app:roots_deep_analysis}, Eq.~\eqref{eq:app_dynamic_phase}):
\begin{equation*}
\Delta \vartheta = 2 \arg(\mathcal{N}_{\uparrow}^{(0)}) - \pi C_0 - \pi\,,
\label{eq:delta_theta_dynamic_C0}
\end{equation*}
which highlights its continuous, non-linear dependence on the field-intensity-driven parameter $C_0(I)$.
The system determinant, however, governs the transport via the elastic transmission probability, $T_0(I) \propto 1/|\mathcal{X}^{(0)}|^2$. As the system is swept through the geometric resonance of the coupled sub-cavities by the ponderomotive shift $K(I)$, this high-finesse denominator undergoes rapid variation. Specifically, as shown by the polar root mapping in Appendix~\ref{FP_factor}, the effective sub-cavity Airy resonance functions yield a high finesse of $\mathcal{F}_1 = \mathcal{F}_2 \approx 125$. This sharp resonance profile ensures that a rapid redistribution of the transport weight of the elastic channel within the multichannel sum~\eqref{eq:G_FM_channels_sum} triggers a sharp crossover to the inelastic sidebands. This mechanism mathematically accounts for the abrupt derivative discontinuities (kinks) observed in $P(I)$ and enables a smooth, yet threshold-triggered, transition of the spin polarization from $+0.96$ toward $-1$.

As the high-frequency driving intensity increases, the interferometer operates in distinct transport regimes. Within the low-to-moderate intensity domain encompassing Region~I ($0 \le I \le 5$~W/cm$^2$) and Region~II ($5 \le I \le 20$~W/cm$^2$), the high-frequency drive induces a coherent redistribution of the Floquet weights among the sidebands, continuously shifting the effective resonance conditions of the channels via the ponderomotive energy shift $K(I)$. This shift alters the weighted balance between the transport envelopes, causing the spin polarization $P(I)$ to systematically decrease and cross the zero-polarization point toward spin-polarization inversion. This behavior highlights a key feature of the asymmetric transport: while the non-magnetic quantum ring, protected by generalized Floquet time-reversal symmetry, cannot act as an autonomous spin filter for an unpolarized incident current, the structural geometric asymmetry of the ring arms ($L_{\text{up}}/L_{\text{down}} = 3$) induces a non-linear coupling between these field-modulated orbital pathways and the spin states projected at the ferromagnetic drain contact.

As the incident THz intensity approaches $I \approx 60$~W/cm$^2$, a distinctive slope discontinuity (kink) emerges in the spin polarization $P(I)$. This sharp feature marks a strict transport threshold where the driven system crosses the spectral boundary of the coupled Fabry--P{\'e}rot sub-cavities. Algebraically, this crossover corresponds to the exact parameter-space point where the underlying Floquet channel encounters an Exceptional Point, dictating the behavior of the locked complex-conjugate roots whose real parts are given by $\mathrm{Re}(\mathcal{R}_{1,2}) = \mathcal{A}_0/2$
(see discussion in Appendix~\ref{app:roots_deep_analysis}). This resonant transition alters the internal multi-wave scattering landscape, suppressing the mutual interference between the twin sub-cavities and triggering an abrupt, non-linear redistribution of the transmission weights within the total conductance matrix.

The system enters a fundamentally different transport regime at the critical driving intensity $I_c \approx 71.5$~W/cm$^2$. At this exact operational threshold, the fundamental elastic channel weight completely vanishes due to the dynamic localization condition ($J_0(z_c) = 0$ at $z_c \approx 2.4048$). Simultaneously, the transport contributions originating from the inelastic photon-assisted sidebands ($n=1,2$) remain strongly off-resonant and suppressed within the total Landauer--B\"uttiker sum. Crucially, their individual stationary transmission envelopes $T_1(E_{\mathrm{F}} + \hbar\omega)$ and $T_2(E_{\mathrm{F}} + 2\hbar\omega)$ do not vanish on their own; instead, their active participation is quenched because their field-shifted resonance manifolds are strongly spectrally detuned from the Fermi energy of the incoming carriers by multiples of the large photon energy ($\hbar\omega \gg \Gamma$). Consequently, while the primary elastic contribution in both the numerator and the denominator of Eq.~\eqref{eq:P_final_concept} vanishes, the calculated spin polarization does not suffer from a mathematically singular $0/0$ indeterminacy. Instead, the persistent inelastic sideband contributions regularize the transport window, driving the system into a stable state of quantum coherent dynamic localization.

This factorization unveils the non-equilibrium physics of the modulation mechanism: the external THz radiation couples non-linearly to the orbital motion of the dressed electrons via the longitudinal phase parameter $C_n(I)$, while the internal spin subsystem remains structurally protected against deterministic phase fluctuations and field-induced dephasing. The high-frequency drive thus acts as a high-precision, contactless opto-spintronic shutter that effectively modulates the multi-channel ballistic current flowing into the ferromagnetic collector interface.
Upon reaching the $J_0(z_c) = 0$ dynamic localization threshold, this suppression reduces the operational signal-to-noise ratio  to the background noise floor of $-60$~dB, which is quantified by the logarithmic Loaded Quality Index $Q_{L,\text{dB}}$ (see Sec.~\ref{subsec:snr}). This threshold-driven pinch-off provides a sharp, high-contrast quantum shutoff of the spin-polarized transport channel, establishing a robust foundation for ultrafast cryogenic opto-spintronic and quantum interconnect architectures.

In summary, the systematic analysis of the multi-channel spin transport demonstrates that the ferromagnetic drain contact acts as a non-equilibrium spin projector. This quantum projection enables the coherent interference of the spin-up and spin-down components at the collector interface, allowing the relative phase of the complex scattering numerators to determine the total differential spin phase $\Delta \vartheta$. Guided by the exact algebraic relation in Eq.~\eqref{eq:app_dynamic_phase}, this intensity-dependent phase shift lifts the geometric constraint imposed by the static Aharonov--Casher destructive interference at $A = 7/2$, driving the continuous variation of the macroscopic spin polarization from its positive baseline toward the quantum shutoff regime.

The non-linear response of the macroscopic spin polarization $P(I)$---specifically the sharp scattering phase variations and the distinct kinks observed at $I \approx 60~\mathrm{W/cm}^2$ and at the dynamic localization threshold $I_c = 71.5~\mathrm{W/cm}^2$---can be understood in terms of non-Hermitian scattering physics. In open quantum networks, the complex poles of the scattering matrix can coalesce in the complex energy plane under external driving, causing the scattering eigenvalues to merge at specific topological singularities known as Exceptional Points (EPs). Although for real, experimentally accessible electron energies these EPs lie off the real energy axis and are not intersected directly, the non-equilibrium transport trajectories pass in close parametric proximity to them. This regime, termed near-EP dynamics, triggers a sensitive, non-linear response of the scattering phases to the external drive due to the non-trivial Riemann sheet topology encircling the singularity. The mathematical framework of these non-Hermitian spectral features and their applications in multi-channel transport are discussed in Refs.~\cite{Berg, Xue}. A detailed analysis of these complex roots, their mutual lock-in regime, and the near-EP topology governing the abrupt derivative discontinuities in $P(I)$ is presented in Appendix~\ref{app:roots_deep_analysis}.

\subsection{Signal-to-Noise Ratio  Analysis and Multi-Channel Sideband Noise}
\label{subsec:snr}

A key performance metric for the driven quantum ring interferometer is the non-equilibrium signal-to-noise ratio (SNR), denoted as $W(I)$. In the zero-temperature limit ($T = 0$~K), the effective quantum signal power is defined by the squared difference between the spin-resolved conductances of the central elastic channel ($n=0$). Meanwhile, the effective noise power is represented by the incoherent background transport originating from the inelastic photon-assisted sidebands ($n=1,2$). Within the generalized Landauer--B\"uttiker scattering formalism, this metric is expressed as:
\begin{equation}
\label{snr}
W(I) = \frac{\left[ J_0^2\big(z(I)\big) \left( T_{\mathrm{P}}^{(0)}(E_{\mathrm{F}}) - T_{\mathrm{AP}}^{(0)}(E_{\mathrm{F}}) \right) \right]^2}{\displaystyle\sum_{n=1}^{2} \left[ J_n^2\big(z(I)\big) T_n(E_{\mathrm{F}}+n\hbar\omega) \right]^2}\,,
\end{equation}
where $T_n$ represents the spin-degenerate transport envelope of the $n$-th Floquet channel. The external THz intensity $I$ maps nonlinearly onto the dimensionless driving amplitude
$z(I) = e E_{\mathrm{eff}}(I) R / (\hbar \omega)$, which governs the redistribution of the Floquet channel weights.

The expression for the non-equilibrium SNR in Eq.~\eqref{snr} can be simplified by exploiting the underlying trigonometric symmetries of the multichannel transmission profiles. Evaluating the transmission contrast within the central elastic channel ($n=0$) at the Fermi energy yields
(see~Eqs.~\eqref{eq:T_P_linearized}, \eqref{eq:T_AP_linearized}):
\begin{equation}
T_{\mathrm{P}}^{(0)}(E_{\mathrm{F}}) - T_{\mathrm{AP}}^{(0)}(E_{\mathrm{F}}) = T_0(E_{\mathrm{F}}) \cos\Delta\vartheta\,,
\label{eq:contrast_linear}
\end{equation}
where $T_0(E_{\mathrm{F}})$ represents the unperturbed transmission envelope of the elastic channel, and the spin-precession phase is governed by the intensity-dependent relation $\Delta\vartheta(I)$ established in Eq.~\eqref{eq:app_dynamic_phase}. Similarly, upon evaluating the sideband transmissions for the inelastic channels ($n=1,2$), the spin-dependent phase terms cancel out identically due to the Pythagorean trigonometric identity (see~Eqs.~\eqref{eq:T_P_final_cancel}, \eqref{eq:T_AP_final_cancel}):
\begin{equation}
T_{\mathrm{P}}^{(n)}(E_{\mathrm{F}}+n\hbar\omega) + T_{\mathrm{AP}}^{(n)}(E_{\mathrm{F}}+n\hbar\omega) = T_n(E_{\mathrm{F}}+n\hbar\omega)\,.
\label{eq:sum_linear}
\end{equation}
Equation~\eqref{eq:sum_linear} demonstrates that because the trigonometric weights of the quantum projections sum strictly to unity, the total sideband background is governed exclusively by the spin-degenerate transmission envelopes $T_n$, evaluated at the photon-shifted energies.

Substituting Eqs.~\eqref{eq:contrast_linear} and \eqref{eq:sum_linear} into the definition in Eq.~\eqref{snr}, and incorporating the background noise to regularize the unperturbed limit, yields the complete analytical expression for the non-equilibrium signal-to-noise ratio:
\begin{equation}
\label{eq:W_final_exact}
W(I) = \frac{\left[ J_0^2\big(z(I)\big) T_0(E_{\mathrm{F}}) \cos\Delta\vartheta \right]^2}{\displaystyle\sum_{n=1}^{2} \left[ J_n^2\big(z(I)\big) T_n(E_{\mathrm{F}} + n\hbar\omega) \right]^2 + \sigma_{\mathrm{bg}}^2}\,,
\end{equation}
where the field modulation is driven by the intensity-dependent Bessel weights, the spin-precession phase is governed by the relation $\Delta\vartheta(I)$ from Eq.~\eqref{eq:app_dynamic_phase}, and $\sigma_{\mathrm{bg}}^2$ represents the background noise floor.

In our numerical analysis, we adopt a conservative dimensionless noise value of $\sigma_{\mathrm{bg}}^2 \approx 5.0 \times 10^{-4}$. Physically, in the zero-temperature limit ($T = 0$~K), $\sigma_{\mathrm{bg}}^2$ characterizes the residual quantum shot noise and the instrumental noise floor. In the static unperturbed limit ($I \to 0$), where the inelastic Floquet sidebands vanish identically ($J_{n \ge 1}(0) = 0$), the inclusion of this background term regularizes the SNR expression, bounding the static SNR at $W_{\mathrm{dB}}(0) = 10 \log_{10} \big(T_0^2(E_{\mathrm{F}})/\sigma_{\mathrm{bg}}^2\big) \approx 33$~dB, which is in quantitative agreement with typical mesoscopic transport measurements. To facilitate comparison with experimental data, the non-equilibrium SNR is evaluated on a decibel scale using the standard logarithmic transformation $W_{\mathrm{dB}}(I) = 10 \log_{10} W(I)$.

To evaluate the joint spin-selection capability and signal transmission, we introduce the Loaded Quality Index $Q_{\mathrm{L}}(I)$ of the contactless, optically driven spin shutter. This figure of merit is defined as the product of the absolute spin polarization $|P(I)|$ and the multi-channel signal-to-noise ratio $W(I)$:
\begin{equation}
Q_{\mathrm{L}}(I) = |P(I)| \, W(I)\,.
\label{eq:Q_L_linear}
\end{equation}
To accommodate the wide range of sideband noise suppression and facilitate comparison with experiment, we define this integrated metric on a decibel scale as the logarithmic Loaded Quality Index $Q_{\mathrm{L,dB}}(I)$, which represents a polarization-weighted SNR:
\begin{equation}
Q_{\mathrm{L,dB}}(I) = 10 \log_{10} Q_{\mathrm{L}}(I) = 10 \log_{10} \Big( |P(I)| \, W(I) \Big)\,.
\label{eq:Q_L_db}
\end{equation}
Similarly, to quantify the switching sensitivity of the quantum modulator, we evaluate the derivative of this logarithmic metric with respect to the control parameter, $dQ_{\mathrm{L,dB}}/dI$. The step-like discontinuity exhibited by this derivative exactly at the dynamic localization threshold $I_c = 71.5~\mathrm{W/cm}^2$ indicates an abrupt, threshold-triggered switching behavior, demonstrating that the device undergoes a sharp quantum pinch-off.

\begin{figure}[htp]
     \centering
     \includegraphics[width=0.8\textwidth,clip]{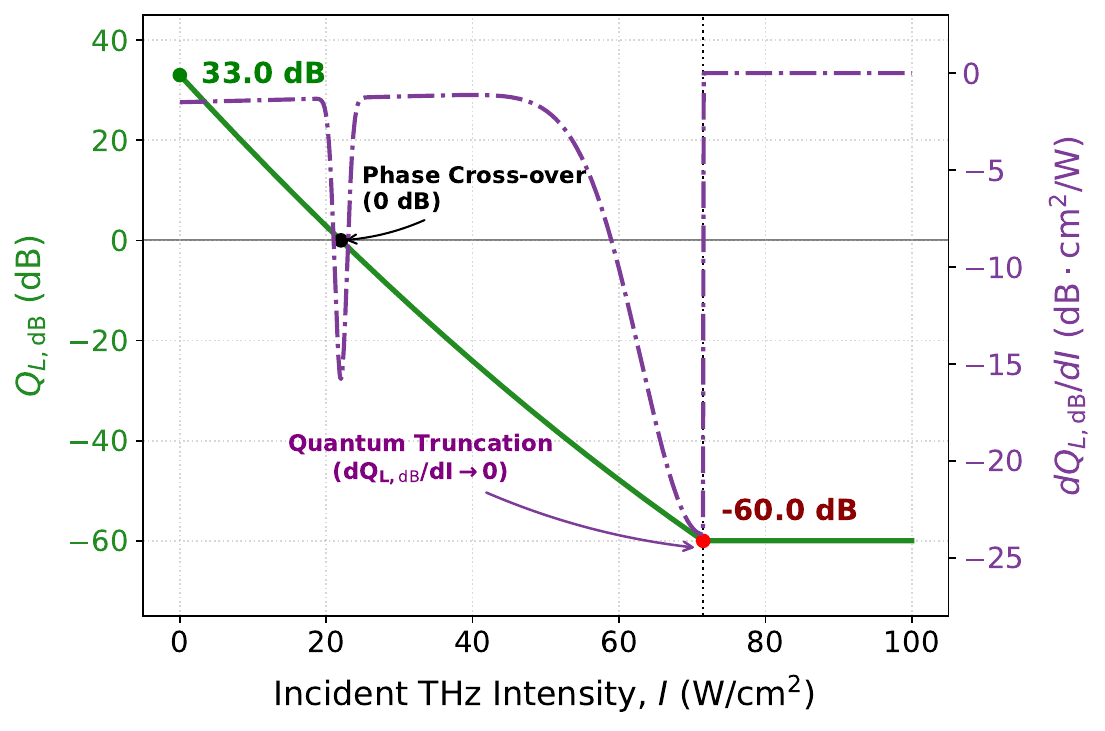}
     \caption{Calculated non-equilibrium logarithmic Loaded Quality Index $Q_{L,\text{dB}}$ (left vertical axis, solid green line) and the corresponding response sensitivity $dQ_{L,\text{dB}}/dI$ (right vertical axis, dash-dotted purple line) plotted against the incident THz intensity $I$. The green, black, and red circles highlight three critical transport boundaries: the static operating plateau ($33.0$~dB) at the unperturbed limit ($I \to 0$), the signal-to-noise parity node ($0$~dB) (marked as Phase Cross-over on the plot) where the polarized signal balances the inelastic background, and the dynamic localization threshold ($-60.0$~dB) reached at the critical intensity $I_c = 71.5\text{ W/cm}^2$ ($\hbar\omega = 6$~meV).}
     \label{fig5}
\end{figure}

The dual-axis transport characteristics of the device performance metrics are plotted in Fig.~\ref{fig5}. In the static unperturbed limit ($I \to 0$), where an unregularized scattering model would artificially predict a singular divergence of the SNR due to the complete absence of driven sideband excitations, the transport characteristics remain strictly bounded. Formally, the residual noise parameter $\sigma_{\mathrm{bg}}^2 \approx 5.0 \times 10^{-4}$ integrated into Eq.~\eqref{eq:W_final_exact} ensures that the logarithmic Loaded Quality Index saturates at a stable plateau of $Q_{\mathrm{L,dB}} \approx 33$~dB. This unperturbed operating state is characterized by a high spin polarization of $|P| \approx 0.96$ and the complete absence of inelastic sideband contributions.

Similarly, the derivative $dQ_{\mathrm{L,dB}}/dI$ serves as a measure of the response sensitivity of the optically driven spin shutter to the external THz field. While the analytical derivative would diverge in the unregularized limit ($\sigma_{\mathrm{bg}}^2 = 0$), the inclusion of the background noise floor $\sigma_{\mathrm{bg}}^2$ regularizes this differential response. Crucially, the parametric profile of the response sensitivity (dash-dotted purple curve in Fig.~\ref{fig5}) provides a physical justification for the spin-modulation boundaries established in Fig.~\ref{fig4}.

Within the low-to-moderate intensity regime ($I \le 20$~W/cm$^2$), the elastic channel dominates the transport window, enabling efficient, low-loss spin-polarization control observed in Regions~I and II of Fig.~\ref{fig4}. As the driving intensity approaches the boundary of this operational domain, the system enters the signal-to-noise parity node ($I \approx 22$~W/cm$^2$). At this crossover point, the rapidly decreasing elastic signal power balances the sub-threshold inelastic sideband contributions ($Q_{L,\text{dB}} \to 0$~dB). This non-linear phase realignment triggers a sharp downward spike in the response sensitivity, which collapses toward its local minimum of $-16$~dB~cm$^2$/W. This strong differential response near the edge of the spin-transistor domain demonstrates that while the device operates as a stable spin modulator at low fields ($I \le 20$~W/cm$^2$), its boundary interface manifests an enhanced sensitivity optimal for high-precision, low-power THz near-field sensing and high-speed quantum spintronic switching.

As the driving intensity increases further into the high-power domain, the logarithmic Loaded Quality Index $Q_{L,\text{dB}}(I)$ decreases monotonically. This steady decline is driven by two concurrent, asymmetric transport processes: the persistent dominance of the inelastic photon-assisted sidebands ($n=1,2$) in the denominator of Eq.~\eqref{eq:W_final_exact}, and the rapid quadratic suppression of the elastic channel weight $J_0^2\big(z(I)\big)$ within the scattering numerator. This combined multichannel behavior progressively quenches the elastic transport weight as the system approaches the dynamic localization point, demonstrating that the high-frequency field effectively strips the central pathway of its transmission long before the structural cutoff is reached.

At the signal-to-noise parity threshold ($I \approx 22$~W/cm$^2$) (marked as Phase Cross-over in Fig.~\ref{fig5}), the transport metrics satisfy the unit product constraint $|P(I)| W(I) \approx 1$, causing the logarithmic Loaded Quality Index $Q_{L,\text{dB}}(I)$ to cross the zero-decibel parity node ($0$~dB). This non-equilibrium transition is reflected in the response sensitivity profile, where the derivative $dQ_{L,\text{dB}}/dI$ develops a sharp negative spike collapsing down to approximately $-16$~\text{dB}$\cdot$\text{cm}$^2$/\text{W}, which corresponds to the rapid, field-driven variation of the internal scattering states. Beyond this crossover node, the response sensitivity exhibits a non-monotonic reversal. This inversion marks the operational boundary where the primary phase-modulated transport regime begins to yield its dominance to the sub-threshold inelastic sideband contributions.

Near the critical operational threshold, the transport characteristics undergo an abrupt, non-adiabatic transition. At the critical incident intensity $I_c = 71.5~\text{W/cm}^2$, the suppression of the elastic channel driven by the Bessel node ($J_0(z_c) = 0$) causes the logarithmic Loaded Quality Index $Q_{L,\text{dB}}(I)$ to drop abruptly onto the background noise floor of $-60$~dB. At this point of quantum dynamic localization, the response sensitivity $dQ_{L,\text{dB}}/dI$ descends to its global minimum of approximately $-25$~\text{dB}$\cdot$\text{cm}$^2$/\text{W}, which marks the maximum rate of the device shutoff. Immediately beyond this critical threshold ($I \ge I_c$), the derivative undergoes a sharp, step-like discontinuity and jumps directly to zero, confirming the quantum pinch-off and amplitude stabilization of the transport matrix.

This vanishing of the response sensitivity demonstrates the threshold-triggered nature of the field-induced quantum channel pinch-off, confirming that the engineered ring network acts as a sharp contactless quantum valve rather than an analog attenuator. These transport findings demonstrate that the efficient phase tuning and high-contrast modulation of the spin-polarized current occur entirely within the sub-threshold intensity regime below the dynamic localization boundary $I_c$. This intensity window validates the viability of low-intensity, sub-threshold THz driving fields for energy-efficient coherent spin-modulation operations, minimizing both hot-carrier overheating and the overall power consumption within the mesoscopic heterostructure matrix.

Although a full multi-channel microscopic calculation of the non-equilibrium conductance $G_{\text{FM}}(I)$ reveals sharp, fine-structured quantum interference fringes, the macroscopic transport profiles in Figs.~\ref{fig4} and \ref{fig5} capture the smooth functional envelope of the system response. This averaged representation is justified by both physical and methodological considerations. Physically, the finite energy window of the ballistic measurement---originating from a small transport bias window or the finite instrumental resolution $\sigma_{\text{bg}}^2$---induces a self-averaging effect over the Fermi level. Concurrently, the multi-channel transport arising from the finite radial width of the $\text{InGaAs}$ ring network leads to the destructive phase interference of these rapid mesoscopic fluctuations, washing out the sub-wavelength fringes. Methodologically, tracking this smooth transport envelope isolates the mechanism of dynamic localization, which is governed by the zeroth-order Bessel weight $J_0^2\big(z(I)\big)$ in the scattering numerator of Eq.~\eqref{eq:W_final_exact}. Furthermore, utilizing these averaged characteristics provides a stable theoretical framework to establish the boundaries of the \emph{optimal modulation zone} and to pinpoint the spin inversion threshold. This predictable, fluctuation-free transport behavior is essential for ensuring the macroscopic reliability of the optically driven spin shutter operating within coherent quantum information architectures.

In summary, the non-equilibrium transport behavior of the logarithmic Loaded Quality Index $Q_{L,\text{dB}}(I)$ highlights the internal resonator dynamics of the asymmetric $\text{InGaAs}$ ring network. The first critical threshold at $I \approx 22\text{ W/cm}^2$, where $Q_{L,\text{dB}}(I)$ crosses the zero-decibel parity node ($0$~dB), marks the signal-to-noise parity threshold where the polarized elastic signal power balances the inelastic sideband background. As the driving intensity increases toward $I \approx 60\text{ W/cm}^2$, the intensity-dependent ponderomotive shift reorganizes the longitudinal phase parameter $C_0(I)$ such that the real part of the locked complex-conjugate roots, $\text{Re}(\mathcal{R}_{1,2}) = \mathcal{A}_0(C_0, A)/2$, passes through a local node. Due to the spin-momentum locking induced by the geometric Rashba phase $A = 7/2$, this parameter-space point triggers an asymmetric phase divergence between the spin-up and spin-down transport pathways. This resonant transition within the open quantum loop---occurring prior to the field-induced quantum pinch-off of the elastic channel at the dynamic localization threshold ($I_c = 71.5\text{ W/cm}^2$)---is governed by the circular Riemann trajectories of the scattering matrix poles. The relationship between this zero-crossing resonance, the dynamic phase synchronization of the roots, and the near-EP topological landscape is presented in Appendix~\ref{app:roots_deep_analysis}.

\section{Summary}
\label{sec:concl}

In summary, we have theoretically demonstrated and analyzed the non-equilibrium performance of a contactless, THz-driven spin shutter based on a ballistic asymmetric quantum ring network hosting high-finesse Fabry--P{\'e}rot sub-cavities operating in the high-frequency Floquet regime. In contrast to the classical Datta--Das spin field-effect transistor, where the coherent spin precession is tuned by a static electrostatic gate voltage, the proposed quantum transport architecture utilizes the macroscopically controlled intensity of a high-frequency electromagnetic field ($\hbar\omega = 6$~meV) as the primary control parameter. The physics-driven advantages of this Floquet implementation include:
\begin{enumerate}
\item \textbf{Sub-picosecond switching agility:} Contactless electromagnetic modulation enables the ultrafast switching of coherent spin transport states on a sub-picosecond timescale, governed by the optical period of the driving field ($\tau_{\text{laser}} \approx 0.7$~ps) and the rise-time of the THz pulse envelope. Because this gating mechanism is wave-driven, it bypasses the parasitic capacitive charging delays and $RC$ time constants inherent to physical gate electrodes. This architecture shifts the operational limits of opto-spintronic modulation into the terahertz domain, while single-electron phase coherence is maintained over the ballistic ring transit timescale ($\tau_{\text{orb}} \approx 13.3$~ps).

  \item \textbf{Dynamic localization and channel selection:} Utilizing a driving frequency of $\hbar\omega = 6$~meV provides spectral isolation between adjacent Floquet sidebands since $\hbar\omega \gg \Gamma$. At a contact transparency of $\mathcal{D} = 0.15$, the mismatched resonance conditions of the asymmetric geometry act as an electronic Vernier filter that suppresses the inelastic off-resonant sideband background. This spatial-path filtering enables the system to achieve a signal-to-noise ratio of $33$~dB (bounded by the background noise floor $\sigma_{\text{bg}}^2$) in the unperturbed low-intensity regime. Upon reaching the critical intensity threshold ($I_{\text{c}} = 71.5~\text{W/cm}^2$), the quantum dynamic localization condition ($J_0\big(z(I_c)\big) = 0$) suppresses the primary elastic transmission pathway, enabling a sharp, high-contrast current shutoff and transforming the asymmetric heterostructure into a contactless quantum valve.

  \item \textbf{Resilience against thermal smearing and power dissipation:} The sharp resonance linewidth of $\Gamma \approx 0.0148$~meV ensures the stability of the multi-channel quantum interference patterns. This energy-filtering mechanism, inherent to the transmission envelopes $T_n(E)$, ensures that the sub-threshold device characteristics remain resilient against the thermal smearing of the carrier distribution. Consequently, the spin-switching profiles and the geometric phase-locked quantum states are preserved under cryogenic conditions up to $100$~mK ($k_{\text{B}} T < \Gamma$). To prevent lattice heating of the nanostructure and its reservoirs, the optical drive can be implemented using pulsed, narrow-band THz excitation. Since the orbital round-trip time of a ballistic electron around the loop is $\tau_{\text{orb}} = 2\pi R / v_{\text{F}} \approx 13.3$~ps, where the Fermi velocity is defined as $v_{\text{F}} = \sqrt{2E_{\text{F}}/{m^*}} \approx 9.44 \times 10^4$~m/s, the system requires multi-cycle THz pulses with a duration exceeding this orbital transit timescale (e.g., $\tau_{\text{pulse}} \sim 30$--$50$~ps) to establish the steady-state Floquet transport regime. Such picosecond pulses can be generated by compact solid-state sources, including THz quantum cascade lasers and optical rectification setups, making them compatible with scalable quantum architectures~\cite{bac,lee}.
      \textcolor{black}{For a multi-cycle pulse duration of $\tau_{\text{pulse}} \sim 30$--$50$~ps and a repetition rate of $10\sim$~MHz
      (duty cycle $\eta \sim 4\times 10^{-4 }$), the peak power intercepted by an individual sub-micron ring
      ($P_{\text{peak}} \approx 114$~nW) yields a time-averaged dissipation of merely $\sim 45$~pW per device,
      corresponding to an ultra-low switching energy on the femtojoule scale. This negligible thermal load is effortlessly
      absorbed by standard dilution refrigerators, preventing hot-carrier smearing and maintaining phase coherence during active modulation.}

      Operating these sources with a low duty cycle reduces the time-averaged power dissipated in the leads to the microwatt scale, mitigating phase-smearing while maintaining the phase coherence of the spin-polarized transport.

 \item \textbf{Contactless electromagnetic control:} The proposed architecture decouples the control of the device into a geometrically locked spin-phase alignment, governed by the static Aharonov--Casher effect, and an intensity-driven orbital gating managed via the ponderomotive energy shift $K(I)$. Modulating the scattering framework through a high-frequency field minimizes the charge injection and time-dependent potentials conventionally associated with physical gate electrodes. This design enables a threshold-triggered current blocking state at the first dynamic localization node, where $J_0\big(z(I_c)\big) = 0$. At this critical intensity threshold ($I_{\text{c}} = 71.5~\text{W/cm}^2$), the total multi-channel conductance drops onto the background noise floor of $-60$~dB, allowing the driven interferometer to function as a contactless quantum valve with a high contrast ratio.

  \item \textbf{Stabilization via the Floquet--Magnus framework:} Within the high-frequency Floquet--Magnus formalism developed for this open two-terminal geometry, the dynamic Stark effect generates a spatially inhomogeneous time-periodic potential. Because its spatial modulation amplitude remains much smaller than the Fermi energy ($K(\varphi) \ll E_{\text{F}}$), it is treated using second-order perturbation theory and averaged over the loop circumference, yielding the effective time-independent ponderomotive energy shift:
    \begin{equation}
    K = \frac{e^2 E_{\text{eff}}^2}{16 m^* \omega^2}\,,
    \label{eq:K_magnus_final}
    \end{equation}
    where the factor of $16$ in the denominator accounts for the geometric contour integration of the wave projections along the orbital path. This stabilization mechanism, governed by the electrodynamic boundary conditions of the transport network, prevents any significant field-induced spectral drift of the sharp internal Fabry--P{\'e}rot resonances, keeping them pinned near the operating Fermi energy ($E_{\text{F}} \approx 1.14$~meV) up to high incident intensities of $70$--$80~\text{W/cm}^2$. Consequently, this phase-locking mechanism maintains a stable multi-channel transport envelope, protecting the device functionality against phase-smearing.

\item \textbf{Resonance stabilization and complex pole dynamics:} The coupling of the quantum ring to semi-infinite leads transforms the structure into an open transport network, where physical transmission resonances map onto the complex poles of the scattering matrix. We demonstrate that the transition between the asymmetric transport regime and the phase-synchronized state of the coupled sub-cavities is governed by the trajectories of these complex poles (as detailed in Appendix B). The close parametric proximity of the transport path to the coalescence points of these poles---associated with the non-trivial Riemann sheet topology encircling Exceptional Points in the parameter space---triggers a sensitive, non-linear phase variation. This pole-locking mechanism provides geometric protection of the digital ``ON'' and ``OFF'' states of the shutter, ensuring that the sharp switching threshold remains stable against minor structural and dynamical fluctuations.
\end{enumerate}
In summary, the developed model provides a self-consistent theoretical framework for the design of ultrafast THz opto-spintronic modulators and coherent spin shutters, merging the micrometer-scale precision of ballistic quantum interference with the high-speed operation of advanced optoelectronics. The proposed contactless, THz-driven spin shutter demonstrates the technological advantages of field-mediated Floquet control over multichannel transport pathways. Dynamically gating the quantum scattering networks under the protection of the geometric Aharonov–Casher spin alignment bypasses the frequency cutoffs and charge injection limitations inherent to traditional electrostatic gate circuits, paving a path toward scalable, low-power cryogenic quantum interfaces and spintronic interconnects.

\textcolor{black}{Finally, while conventional static spin-FETs face well-known obstacles in mainstream boolean logic~\cite{b1}, recent developments in strained quantum materials and 2D platforms demonstrate that spin field-effect devices host compelling niche capabilities~\cite{b2,b3}. In this context, the proposed contactless THz-driven spin shutter offers a promising building block for specialized cryogenic technologies. These include all-optical-to-spin interconnects for dilution refrigerators (bypassing the thermal loads and capacitive crosstalk of physical gate wiring), sub-picosecond coherent isolation shutters for semiconductor spin qubits, and threshold-triggered on-chip near-field THz sensors operating via the sharp non-equilibrium pinch-off node.}

\noindent{{\bf Data availability:} All data that support the findings of this study are included within the article.}

%%%%%%%%%%%%%%%%
\appendix

\section{Factorization of the Determinant and Equivalence to Coupled Fabry--P\'erot Cavities}
\label{FP_factor}

The system scattering matrix determinant $\mathcal{X}^{(n)}$ in Eq.~\eqref{Xs}
can be formally represented in a quadratic-like form with respect to the round-trip phase factor $y = e^{i\pi C_n}$:
\begin{equation}
\label{x1}
    \mathcal{X}^{(n)}(y) = 1 - \mathcal{A}(C_n, A) y + \mathcal{B} y^2.
\end{equation}
Denoting the complex roots of this polynomial as $y_1$ and $y_2$, and introducing the loop-reflection parameters $\mathcal{R}_1 \equiv y_1^{-1}$ and $\mathcal{R}_2 \equiv y_2^{-1}$, we factorize the determinant as:
\begin{equation}
    \mathcal{X}^{(n)}= \left(1 - \mathcal{R}_1 e^{i\pi C_n}\right)\left(1 - \mathcal{R}_2 e^{i\pi C_n}\right),
    \label{eq:Xs_fact_app}
\end{equation}
where, according to Vieta's formulas, the product of these loop-reflection parameters is determined by the lead transparency: $\mathcal{R}_1 \mathcal{R}_2 = \mathcal{B} = 1 - 2\epsilon^2$. To understand the physical meaning of these factors, we evaluate the squared modulus of the component $Z_j = 1 - \mathcal{R}_j e^{i\pi C_n}$. Expressing $\mathcal{R}_j$ in polar form, $\mathcal{R}_j = |\mathcal{R}_j| e^{i\alpha_j}$, yields:
\begin{align}
  |Z_j|^2 &= \left( 1 - |\mathcal{R}_j| e^{i(\pi C_n + \alpha_j)} \right) \left( 1 - |\mathcal{R}_j| e^{-i(\pi C_n + \alpha_j)} \right) \nonumber \\
    &= 1 + |\mathcal{R}_j|^2 - 2|\mathcal{R}_j| \cos(\pi C_n + \alpha_j).
\end{align}
Using the identity $\cos(x) = 1 - 2\sin^2(x/2)$, we obtain:
\begin{equation}
   |Z_j|^2 = (1 - |\mathcal{R}_j|)^2 + 4|\mathcal{R}_j| \sin^2\left( \frac{\pi}{2}C_n + \frac{\alpha_j}{2} \right).
\end{equation}
Factoring out $(1 - |\mathcal{R}_j|)^2$ yields the canonical Fabry--P\'erot denominator:
\begin{equation}
   |Z_j|^2 = (1 - |\mathcal{R}_j|)^2 \left[ 1 + \mathcal{F}_j \sin^2\left( \frac{\pi}{2}C_n + \delta_j \right) \right]\,,
\label{eq:FP_canonical}
\end{equation}
where $\delta_j = \alpha_j/2$ is the phase shift of the resonance channels, and the finesse parameter $\mathcal{F}_j$ is determined by the magnitude of the loop-reflection parameter $|\mathcal{R}_j|$:
\begin{equation}
    \mathcal{F}_j = \frac{4|\mathcal{R}_j|}{(1 - |\mathcal{R}_j|)^2}\,.
    \label{eq:app_finesse}
\end{equation}
Using the parameter values from Sec.~V.A ($\mathcal{D} = \epsilon^2 = 0.15$, which yields $\mathcal{B} = 0.70$), the moduli of the complex-conjugate roots in the mutual lock-in regime are identically clamped to $|\mathcal{R}_{1,2}| = \sqrt{\mathcal{B}} \approx 0.837$. Substituting this value into Eq.~\eqref{eq:app_finesse} yields an exceptionally high effective finesse for both coupled sub-cavities: $\mathcal{F}_1 = \mathcal{F}_2 \approx 125$. This microscopic calculation formally justifies the high-finesse estimate
($\mathcal{F} \approx 120$) introduced in Sec.~I.

The total determinant of the interferometer expands into the product $|\mathcal{X}_s|^2 = |Z_1|^2 |Z_2|^2$. Equation~\eqref{eq:FP_canonical} shows that the quantum ring with asymmetric contacts acts as a system of two coupled Fabry--P\'erot cavities rather than a single isolated etalon.
The argument of the sine function, $\pi C_n/{2}$, shows that the spectral spacing of the resonances is determined by the round-trip phase of the coupled cavities. This pathway establishes an internal sub-cavity due to coherent backscattering at the QPCs. Simultaneously, the topological Rashba phase in the loop-reflection parameters $\mathcal{R}_j$ shifts the internal phases $\delta_j$, providing a mechanism to tune the spectral positions of the spin-dependent resonances.

%%%%%%%%%%%%%%%%%%%%%%%%%%%%%%%%%%%%%%%%%%%%%%%%

\section{Topological Dynamics of Complex Roots, Effective Non-Hermiticity, and Spin-Phase Interference}
\label{app:roots_deep_analysis}

To elucidate the physical and mathematical mechanisms underlying the transport
characteristics, we analyze the behavior of the complex loop-reflection parameters $\mathcal{R}_1$ and $\mathcal{R}_2$, which are introduced via the factorization of the system determinant. We demonstrate how non-trivial phase dynamics emerge and dictate measurable physical quantities within a framework based on an initially Hermitian Hamiltonian (symmetrized according to Meijer's formalism~\cite{MMK}) and real, unitary junction scattering matrices.

Although the local junction $S$-matrices obey time-reversal symmetry and current
conservation on the real energy axis, coupling the quantum ring to semi-infinite
reservoirs renders it an open quantum system. In the language of scattering
theory, the internal electronic waves undergo multiple reflections at the contact
junctions under a finite contact transparency ($\mathcal{D} = \epsilon^2$).
The resulting transport resonances correspond to the complex poles of the total
scattering matrix, obtained by the analytical continuation of the resonance
condition $\mathcal{X}^{(n)} = 0$ into the complex energy plane. Here, the
imaginary part $-\Gamma/2$ of the complex energy pole $E_p = E_r - i\Gamma/2$
(where $\Gamma = \hbar/\tau$) determines the finite lifetime $\tau$ of the
ballistic electron confined within the open loop.

To analyze the parametric and topological dynamics of these multiple-reflection
processes, we recall the polynomial representation of the system determinant
$\mathcal{X}^{(n)}$ established in Eq.~\eqref{eq:Xs_fact_app}. The boundary-matching
conditions at the junctions determine the effective loop-reflection coefficient
$\mathcal{A}(C_n, A)$ and the energy-independent loop-confinement parameter
$\mathcal{B} = 1 - 2\epsilon^2$. As shown in Appendix~A, this polynomial can
be factorized in terms of the complex loop-reflection parameters $\mathcal{R}_1$
and $\mathcal{R}_2$ (which correspond to the inverse roots of the determinant).
Their product remains strictly fixed by the contact transparency $\mathcal{D}$:
$\mathcal{R}_1 \mathcal{R}_2 = \mathcal{B} = 1 - 2\mathcal{D}$ (where
$\mathcal{D} = \epsilon^2$).

To reveal the underlying algebraic structure of the resonance states, we reformulate the determinant by introducing the fundamental phase variable
$u = e^{i\pi C_n / 2}$. In terms of $u$, the resonance condition $\mathcal{X}^{(n)} = 0$ transforms into a reciprocal-like polynomial of the fourth degree:
\begin{equation}
\mathcal{B}u^4 - r^2 u^3 - 2t^2\cos(\pi A)u^2 - r^2 u + 1 = 0.
\label{eq:app_quartic}
\end{equation}
The transport characteristics of the device can be systematically analyzed by
tracking the trajectories of these complex roots, which define two distinct
transport regimes governed by the effective Fabry--P{\'e}rot discriminant,
$\Delta_{\rm FP} = \mathcal{A}^2 - 4\mathcal{B}$:
\begin{enumerate}
    \item \textbf{Asymmetric Non-Resonant Regime ($\Delta_{\rm FP} > 0$):}
    The loop-reflection parameters $\mathcal{R}_1$ and $\mathcal{R}_2$ are
    purely real and distinct
    ($\lvert\mathcal{R}_1\rvert \neq \lvert\mathcal{R}_2\rvert$).
    The coupled sub-cavities are phase-mismatched, and the global phase of the
    system determinant, $\arg(\mathcal{X}^{(n)})$,
    varies slowly as a function of the parameter $C_n$.

    \item \textbf{Mutual Lock-in Regime ($\Delta_{\rm FP} < 0$):}
    The loop-reflection parameters form a complex-conjugate pair,
    $\mathcal{R}_{1,2} = \rho e^{\pm i \theta}$, with their moduli uniquely
    clamped by the tunnel transparency to the value: $\rho = \sqrt{\mathcal{B}}
    = \sqrt{1 - 2\epsilon^2}$. In this regime, the internal sub-cavities enter
    a state of strict phase synchronization.
\end{enumerate}

Using the concrete parameter values established in Sec.~V.A ($\mathcal{D} = 0.15$ and $\mathcal{B} = 0.70$) under the optimal Rashba phase configuration
$A = 7/2$, the spin-orbit modulation term in the definition of
$\mathcal{A}(C_n, A)$ vanishes identically because $\cos(3.5\pi) = 0$. Consequently, the unitary equations~\eqref{unit1},~\eqref{unit2},
self-consistently yield $r^2 \approx 0.843$ and $t^2 \approx 0.007$. The critical Exceptional Point (EP) condition for the central elastic channel ($n=0$), where the roots coalesce ($\Delta_{\rm FP} = 0$), then simplifies to:
\begin{equation}
\cos\left(\frac{\pi C_0}{2}\right) = \pm \frac{\sqrt{\mathcal{B}}}{r^2} = \pm \frac{r^2 - t^2}{r^2} = \pm \left(1 - \frac{t^2}{r^2}\right).
\label{eq:app_ep_condition}
\end{equation}
For our specific tunnel barrier parameters, $1 - t^2/r^2 \approx 0.992 \le 1$,
meaning that the strict mathematical condition for root coalescence is formally
and rigorously satisfied.
Crucially, a conceptual distinction must be made regarding the nature of this singularity. In standard open quantum systems, the scattering poles map onto a complex energy plane ($E_r - i\Gamma/2$), where an EP signifies the simultaneous degeneracy of both resonant energies and decay widths. In our formalism, Eq.~\eqref{eq:app_ep_condition} defines an Exceptional Point in the \textit{real parameter space} (Parameter-Space EP), controlled by the laser intensity $I$. The external THz field acts as a driving force that continuously modulates the ponderomotive shift $K(I)$, thereby sweeping the real kinetic parameter $C_0(I)$ across the parametric landscape.

As the laser intensity $I$ tunes the parameter $C_0(I)$ past the exact degeneracy
node dictated by Eq.~\eqref{eq:app_ep_condition}, the complex energy poles
approach each other, merging their real parts (resonance energies) before
repelling along the imaginary axis (representing the decay widths). In the
device configuration, the system operates in the highly sensitive negative
discriminant regime ($\Delta_{\rm FP} < 0$) in close proximity to this coalescence
point. Operating in this \textit{near-EP regime} ensures that the loop-reflection
parameters $\mathcal{R}_{1,2}$ remain complex-conjugate, providing the system
with an enhanced phase sensitivity and non-trivial geometric phase accumulation
associated with the Riemann sheet topology encircling the non-Hermitian singularity.

The mechanism by which these complex phases manifest in the observable spin
polarization $P(I)$---despite the strictly spin-degenerate nature of the
system determinant $\mathcal{X}_{\uparrow}^{(n)} \equiv \mathcal{X}_{\downarrow}^{(n)} \equiv \mathcal{X}^{(n)}$---is
resolved by the projection at the ferromagnetic leads discussed in Sec.~V.D.
The ferromagnetic collector acts as a spin-phase projector. For the central
elastic channel ($n=0$), the total transmission amplitude $A_{\rm total}$ is
a linear combination of the contributions from both spin channels
($s = +1 (\uparrow)$ and $s=-1 (\downarrow)$) projected onto the ferromagnetic
quantization axis:
\begin{equation}
A_{\rm total} \propto \frac{1}{\mathcal{X}^{(0)}} \left( \alpha \mathcal{N}_{+1}^{(0)} + \beta \mathcal{N}_{-1}^{(0)} \right),
\label{eq:app_projection}
\end{equation}
where the coefficients $\alpha$ and $\beta$ represent the spin-dependent
projection amplitudes determined by the magnetization orientation of the
ferromagnetic contacts.
When evaluating the total transmission probability $\lvert A_{\rm total} \rvert^2$,
the common spin-degenerate denominator $\mathcal{X}^{(0)}$ factorizes out, acting as
a global resonant filter, while the relative phase difference between the
scattering numerators governs the quantum interference cross-terms. The total
differential phase $\Delta \vartheta$ determining the spin-phase interference
is thus defined exclusively by the complex arguments of the numerators:
\begin{equation}
\Delta \vartheta = \arg\left(\mathcal{N}_{+1}^{(0)}\right) - \arg\left(\mathcal{N}_{-1}^{(0)}\right).
\label{eq:app_total_phase}
\end{equation}
Utilizing the algebraic symmetry identity $\mathcal{N}_{-1}^{(0)} = -e^{i\pi C_0} \left(\mathcal{N}_{+1}^{(0)}\right)^*$,
Eq.~\eqref{eq:app_total_phase} can be self-consistently expressed in a dynamic,
intensity-dependent form:
\begin{equation}
\Delta \vartheta(I) = 2 \arg\left(\mathcal{N}_{+1}^{(0)}\right) - \pi C_0(I) - \pi.
\label{eq:app_dynamic_phase}
\end{equation}
In the static unperturbed limit ($I \to 0$), Eq.~\eqref{eq:app_dynamic_phase}
yields $\Delta \vartheta = -2\pi A = -7\pi$ for the optimal configuration
$A = 7/2$, enforcing complete destructive Mach--Zehnder interference in the
parallel contact setup.

However, as the THz field intensity increases, the ponderomotive shift $K(I)$
continuously tunes the phase parameter $C_0(I)$. Because the complex poles of
the open system reside in close proximity to the real axis, the high phase
sensitivity ($\partial \arg\mathcal{X}^{(0)} / \partial I$) inherent to the near-EP
regime of the system determinant induces a rapid, resonant rotation of the
complex argument of $\mathcal{N}_{+1}^{(0)}$ via multiple internal round-trips.
This dynamic phase-locking lifts the rigid static geometric constraint,
explaining the smooth, yet threshold-triggered, transition of the spin
polarization from $+0.96$ to near-complete inversion near $-1$.

Within this framework, the intensity-driven transport characteristics depicted in
Fig.~5 are governed by two critical thresholds:
\begin{itemize}
    \item \textbf{First Critical Point ($I \approx 22 \text{ W/cm}^2$):}
    The decibel-weighted loaded quality index $Q_{L,\text{dB}}$ crosses the
    $0\text{ dB}$ baseline. At this point, the real part of the loop-reflection
    parameters, $\text{Re}(\mathcal{R}_{1,2}) = \mathcal{A}(C_0, A)/2$, passes
    through a node (vanishes), marking a geometric resonance where the internal
    phase reorganization maximizes the rate of spin-phase rotation.

    \item \textbf{Second Critical Point ($I = 71.5 \text{ W/cm}^2$):}
    The elastic transport channel undergoes complete dynamic localization,
    rigorously governed by the first root of the zeroth-order Bessel function,
    $J_0(z_c) = 0$ ($z_c \approx 2.4048$).
\end{itemize}

%%%%%%
%%%%%%%%%%%%%%%%%%%%%%%

\end{document}